\documentclass[openany]{nature}
\usepackage{graphicx}
\usepackage{float}
\usepackage{verbatim}
\usepackage{amsmath}
\usepackage{amssymb}
\usepackage{subfigure}
\usepackage{cases}
\usepackage{mathrsfs}
\usepackage[flushleft]{threeparttable}
\usepackage{amssymb}
\usepackage{pifont}
\usepackage{epstopdf}
\usepackage{xcolor}
\usepackage{hyperref}
\usepackage[all]{hypcap}
\usepackage{mwe,tikz}
\usepackage{bm}
\usepackage{multirow}
\usepackage{longtable}
\usepackage{xspace}
\usepackage{rotating}
\usepackage{CJK}
\usepackage{url} 
\usepackage{upgreek}
\usepackage{booktabs}
\usepackage{textcomp}
\usepackage{makecell}
\usepackage{enumitem}
\usepackage{csquotes}
\usepackage{soul}
\usepackage{tabularx}
\usepackage{overpic}
\usepackage{caption}
\usepackage{float}
\usepackage{threeparttable}
\usepackage{threeparttablex}
\usepackage{subfiles}
\usepackage{array}

\makeatletter
 
\def\emph#1 {\textit{ #1 } }

\let\saved@includegraphics\includegraphics
\AtBeginDocument{\let\includegraphics\saved@includegraphics}
\renewenvironment*{figure}{\@float{figure}}{\end@float}

\makeatother

\newcommand{\feii}{\mbox{Fe\,{\sc ii}}}
\newcommand{\mgii}{\mbox{Mg\,{\sc ii}}}
\newcommand{\siii}{\mbox{Si\,{\sc ii}}}

\title{An extremely soft and weak fast X-ray transient associated with a luminous supernova}

\author{W.-X. Li$^{1}$\thanks{These authors contributed equally to this work}, Z.-P. Zhu$^{1*}$, X.-Z. Zou$^{2*}$, J.-J. Geng$^{3*}$, L.-D. Liu$^{4}$, Y.-H. Wang$^{5,6}$, R.-Z. Li$^{7}$,
D. Xu$^{1}$\thanks{E-mail: dxu@nao.cas.cn}, 
H. Sun$^{1}$\thanks{E-mail: hsun@nao.cas.cn},
X.-F. Wang$^{8}$\thanks{E-mail: wang\_xf@mail.tsinghua.edu.cn}, 
Y.-W. Yu$^{4}$\thanks{E-mail: yuyw@ccnu.edu.cn}
% ...
}

\begin{document}
\captionsetup[table]{name={\bf Table}}
\captionsetup[figure]{name={\bf Fig.}}

\maketitle

\begin{affiliations}
\item { National Astronomical Observatories, Chinese Academy of Sciences, Beijing 100101, China }
\item { South-Western Institute for Astronomy Research, Yunnan University, Kunming, Yunnan 650504, People's Republic of China }
\item { Purple Mountain Observatory, Chinese Academy of Sciences, Nanjing 210023, China }
\item { Institute of Astrophysics, Central China Normal University, Wuhan 430079, China }
\item { Nevada Center for Astrophysics, University of Nevada, 4505 S. Maryland Pkwy., Las Vegas, NV 89154-4002, USA }
\item { Department of Physics and Astronomy, University of Nevada, 4505 S. Maryland Pkwy., Las Vegas, NV 89154-4002, USA }
\item { Yunnan Observatories, Chinese Academy of Sciences, Kunming 650216, People's Republic of China }
\item { Physics Department, Tsinghua University, Beijing, 100084, China }
\item { Department of Astronomy, University of California, Berkeley, CA 94720-3411, USA }
\item { Hagler Institute for Advanced Study, Texas A\&M University, 3572 TAMU, College Station, TX 77843, USA }
\item { School of Astronomy and Space Science, University of Chinese Academy of Sciences, Chinese Academy of Sciences, Beijing 100049, China }
\item { Instituto de Astrof\'{\i}sica de Canarias, V\'{\i}a L\'actea, 38205 La Laguna, Tenerife, Spain }
\item { Universidad de La Laguna, Departamento de Astrof\'{\i}sica, 38206 La Laguna, Tenerife, Spain }
\item { Shanghai Astronomical Observatory, Chinese Academy of Sciences, 80 Nandan Road, Shanghai 200030, China }
\item { Key Laboratory of Radio Astronomy and Technology, Chinese Academy of Sciences, A20 Datun Road, Chaoyang District, Beijing 100101, China }
\item { South African Astronomical Observatory, PO Box 9, Observatory 7935, Cape Town, South Africa }
\item { Department of Astronomy, University of Cape Town, Private Bag X3, Rondebosch 7701, Cape Town, South Africa }
\item { Instituto de Astrof\'isica de Andaluc\'ia (IAA-CSIC), Glorieta de la Astronom\'ia s/n, 18008 Granada, Spain }
\item { Ingeniería de Sistemas y Autom\'atica, Universidad de M\'alaga, Unidad Asociada al CSIC por el IAA, Escuela de Ingenier\'ias Industriales, Arquitecto Francisco Pe\~nalosa, 6, Campanillas, 29071 M\'alaga, Spain }
\item { Department of Astronomy, School of Physics, Huazhong University of Science and Technology, Wuhan, 430074, People’s Republic of China }
\item { Niels Bohr Institute, University of Copenhagen, Jagtvej 155, DK-2200, Copenhagen N, Denmark }
\item { The Cosmic Dawn Centre (DAWN), Denmark }
\item { Las Cumbres Observatory, 6740 Cortona Drive, Suite 102, Goleta, CA 93117-5575, USA }
\item { Department of Physics, University of California, Santa Barbara, CA 93106-9530, USA }
\item { Department of Physics, Royal Holloway - University of London, Egham, TW20 0EX, U.K. }
\item { Observatories of the Carnegie Institution for Science, 813 Santa Barbara St, Pasadena, CA 91101, USA }
\item { Special Astrophysical Observatory of the Russian Academy of Sciences, Nizhny Arkhyz, 369167, Russia }
\item { Centro Brasileiro de Pesquisas F\'isicas, Rua Dr. Xavier Sigaud 150, 22290-180 Rio de Janeiro, RJ, Brazil }
\item { Key Laboratory of Particle Astrophysics, Institute  of High Energy Physics, Chinese Academy of Sciences, Beijing 100049, China }
\item { Centre for Astrophysics and Supercomputing, Swinburne University of Technology, Hawthorn, VIC 3122, Australia }
\item { ARC Centre of Excellence for Gravitational Wave Discovery (OzGrav), Hawthorn, VIC 3122, Australia }
\item { Department of Physics, The University of Hong Kong, Pokfulam Road, Hong Kong }
\item { Sternberg Astronomical Institute, Moscow State University, Moscow, 119234, Russia }
\item { Faculty of Science, University of Helsinki, Gustaf Hallströmin katu 2, FI-00014 Helsinki, Finland }
\item { Institute of Physics and Technology, Ural Federal University, Mira str. 19, 620002 Ekaterinburg }
\item { Instituto de F\'{i}sica y Astronom\'{i}a, Facultad de Ciencias, Universidad de Valpara\'{i}so, Av. Gran Breta{\~n}a 1111, Valpara\'{i}so, Chile }
\item { Millennium Institute of Astrophysics, Nuncio Monse{\~n}or Sotero Sanz 100, Of. 104, Providencia, Santiago, Chile }
\item { Department of Physics, Florida State University, 77 Chieftan Way, Tallahassee, FL 32306, USA }
\item { Guangxi Key Laboratory for Relativistic Astrophysics, School of Physical Science and Technology, Guangxi University, Nanning 530004 China }
\item { School of Astronomy and Space Sciences, University of Science and Technology of China, 230026, Hefei People's Republic of China }
\item { Cahill Center for Astrophysics, California Institute of Technology, Pasadena, CA 91125, USA }
\item {NASA Hubble Fellow}
\item { Center for Interdisciplinary Exploration and Research in Astrophysics (CIERA), Northwestern University, 1800 Sherman Ave, Evanston, IL 60201, USA }
\item { Space Telescope Science Institute, 3700 San Martin Drive, Baltimore, MD 21218, USA }
\item {Instituto de Alta Investigación, Universidad de Tarapacá, Casilla 7D, Arica, Chile}
\item { Department of Physics, University of Johannesburg, PO Box 524, Auckland Park 2006, South Africa }
\item { Institute for Frontiers in Astronomy and Astrophysics, Beijing Normal University, Beijing, 102206, People’s Republic of China }
\item { Shanghai Institute of Technical Physics, Chinese Academy of Sciences, Shanghai, 200083, China }
\item { Adler Planetarium, 1300 S Dusable Lk Shr Dr, Chicago, IL 60605 }
\item { Chinese Academy of Sciences South America Center for Astronomy (CASSACA), National Astronomical Observatories, CAS, Beijing, China }
\item { Mitchell Institute for Fundamental Physics and Astronomy, Texas A\&M University, College Station, TX, USA }
\item { University of Chinese Academy of Sciences, Beijing 100049, China }
\item { Beijing Planetarium, Beijing Academy of Sciences and Technology, Beijing 100044, China }
\item { Departamento de Astronom\'{i}a, Universidad de Chile, Las Condes, 7591245 Santiago, Chile }
\item { Innovation Academy for Microsatellites, Chinese Academy of Sciences, Shanghai, 201210, China }
\end{affiliations}

\begin{abstract}
Long gamma-ray bursts (LGRBs), including their subclasses of low-luminosity GRBs (LL-GRBs) and X-ray flashes (XRFs) characterized by low spectral peak energies, are known to be associated with broad-lined Type Ic supernovae (SNe Ic-BL), which result from the core collapse of massive stars that lose their outer hydrogen and helium envelopes. However, the soft and weak end of the GRB/XRF population remains largely unexplored, due to the limited sensitivity to soft X-ray emission. Here we report the discovery of a fast X-ray transient, EP250108a, detected by the Einstein Probe (EP) in the soft X-ray band at redshift $z = 0.176$, which was followed up by extensive multiband observations. EP250108a shares similar X-ray luminosity as XRF\,060218, the prototype of XRFs,  but it extends GRBs/XRFs down to the unprecedentedly soft and weak regimes, with its $E_{\rm peak} \lesssim 1.8\,\mathrm{keV}$ and $E_{\rm iso} \lesssim 10^{49}\, \mathrm{erg}$, respectively. Meanwhile, EP250108a is found to be associated with SN\,2025kg, one of the most luminous and possibly magnetar-powered SNe Ic-BL detected so far. Modeling of the well-sampled optical light curves favors a mildly relativistic outflow as the origin of this event. This discovery demonstrates that EP, with its unique capability, is opening a new observational window into the diverse outcomes of death of massive stars.

\end{abstract}

Long-duration gamma-ray bursts (LGRBs), including the subclasses of low-luminosity GRBs (LL-GRBs) and X-ray flashes (XRFs), produce flares of soft gamma-ray (a few hundred $\rm keV$) or hard X-ray (at least a few $\rm keV$ and higher) emission with durations from a few to $> 1000$ seconds\cite{Zhang18}. In contrast, fast X-ray transients (FXTs) produce multiple flashes of soft X-ray (typically $\rm 0.5-4\,keV$) emission with durations from minutes to hours\cite{Jonker2013ApJ,Xue2019,Quirola2022}. A cosmological FXT may manifest itself as a soft X-ray transient with no GRB component\cite{SunH24}, a soft X-ray counterpart of a GRB in its prompt phase\cite{Liu+2025}, or a very early X-ray afterglow of a GRB. The Einstein Probe (EP) is now revolutionizing this field by discovering tens of FXTs yearly with good localizations, enabling multiband rapid follow-up observations.

EP250108a was discovered by the Wide-field X-ray Telescope (WXT) onboard EP in the $0.5$-–$4\,\mathrm{keV}$ band at $T_0 = 12^{\rm hr}47^{\rm m}35^{\rm s}$ on 8 January 2025 (UTC dates are used throughout this paper)\cite{2025GCN.38861....1L}. No significant gamma-ray counterpart was observed (see Methods). Follow-up observations identified an optical counterpart\cite{2025GCN.38878....1E}, though no emission was detected in the soft X-ray or radio bands (see Methods). Spectroscopic classification revealed the optical counterpart to be a broad-lined Type Ic supernova (SN Ic-BL) \cite{2025GCN.38984....1X}, designated SN\,2025kg, at a redshift $z = 0.176$\cite{2025TNSAN..17....1Z} (Methods). 

\begin{figure}[htbp]
\centering
\begin{minipage}[t][-0cm][b]{0.48\textwidth}
    \centering
    \begin{minipage}[b]{\textwidth}
        \centering
        \begin{overpic}[width=0.95\textwidth]{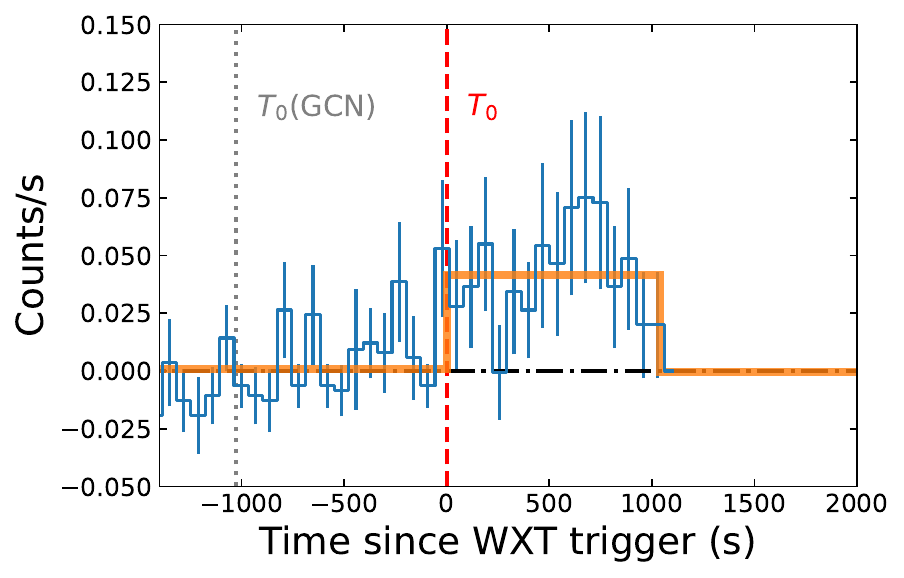}
        \put(-6,66){{\textbf{a}}}
        \end{overpic}
    \end{minipage}
    \begin{minipage}[b]{\textwidth}
        \centering
        \begin{overpic}[width=0.95\textwidth]{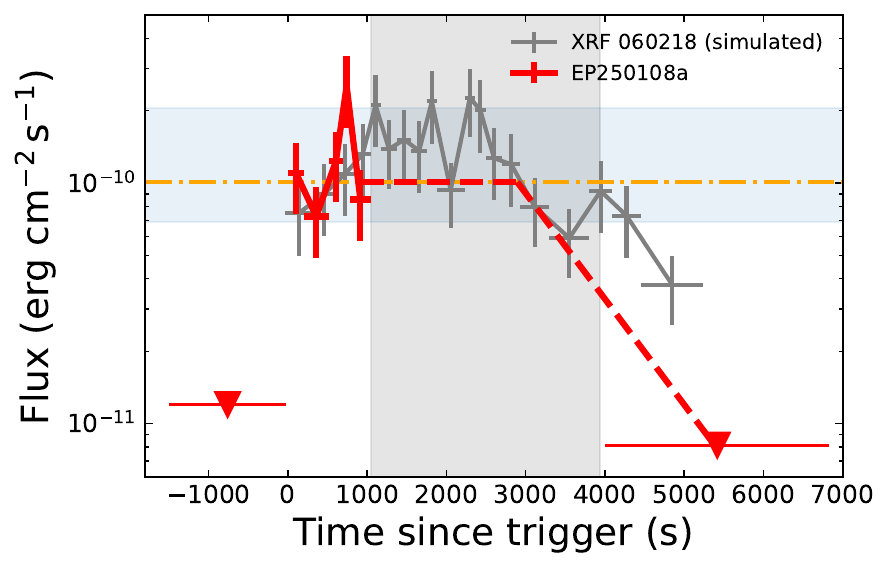}
        \put(-6,68){\textbf{b}}
        \end{overpic}
    \end{minipage}
\end{minipage}
\begin{minipage}[b]{0.5\textwidth}
    \centering
        \begin{overpic}[width=0.98\textwidth]{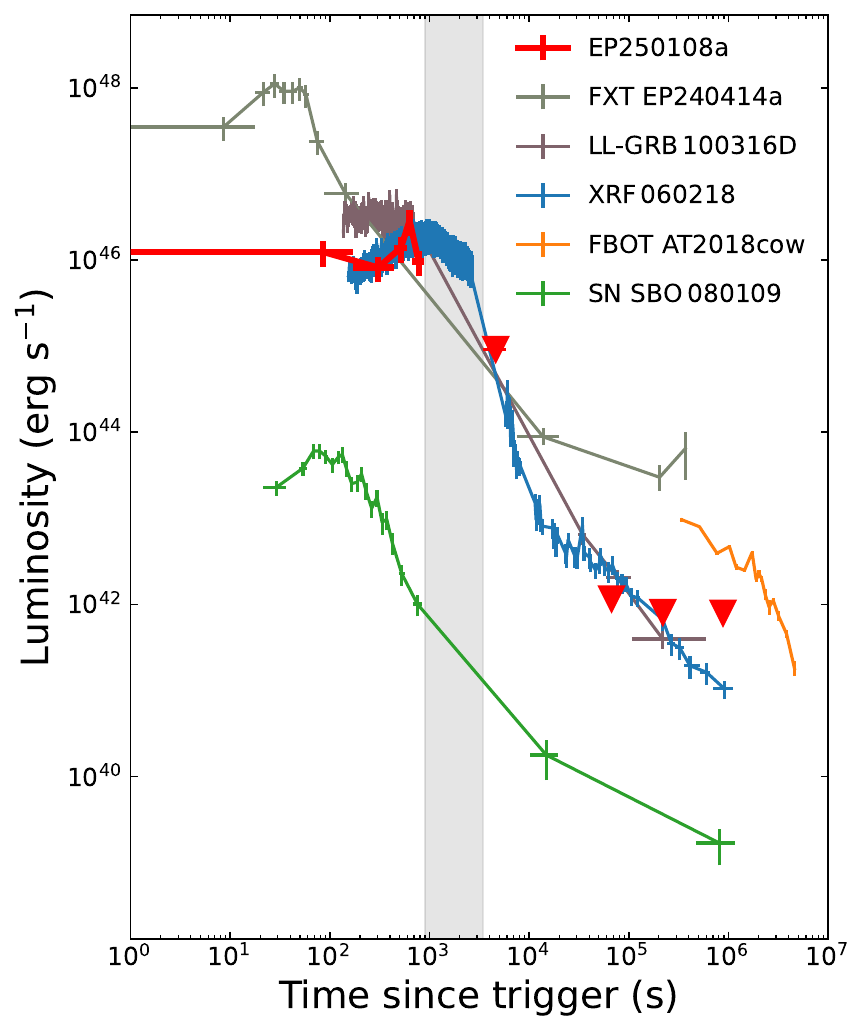}
        \put(0,100){\textbf{c}}
        \end{overpic}
\end{minipage}
\caption{
\noindent\textbf{The X-ray light curve of EP250108a.} 
\textbf{a}, The light curve of the count rate in the 0.5--4\,keV band. A bin size of 70\,s is adopted owing to the relatively low photon counts. The redefined $T_0=$ 2025-01-08 T 12:47:35.72 (red dashed line) is 1027.4\,s later than the GCN report time (grey dotted line). The orange line shows the light curve rebinned using the Bayesian block method, which is used to determine the updated $T_0$ of EP250108a. The black dot-dashed line indicates the zero count-rate level. 
\textbf{b}, The unabsorbed flux of EP250108a and XRF 060218 in the 0.5--4\,keV band. Count rates were rebinned to achieve an SNR of 3$\sigma$, and subsequently converted to unabsorbed fluxes using the best-fit spectral models. The light curve of XRF\,060218 was derived from the \textit{Swift} Burst Analyser\cite{Evans2010}, assuming it was observed by WXT at $z=0.176$ (see Methods). The grey region indicates the period of Earth occultation, and the triangle symbols mark the WXT upper limits of EP250108a. The orange dash-dotted line and the shaded cyan region show the average observed flux and its 1$\sigma$ uncertainties for the WXT data of EP250108a. 
The red dashed line indicates the flux evolution of EP250108a under the assumption that it maintained its average observed flux during part of the Earth-occultation period, and subsequently decayed with the same temporal slope as XRF\,060218, reaching the observed upper limit (see Methods).
\textbf{c}, Rest-frame 0.5--10\,keV X-ray light curve of EP250108a compared with those of representative XRFs, including the fast blue optical transient (FBOT) AT2018cow\cite{2019ApJ...872...18M}, the supernova shock breakout (SBO) event 080109\cite{Modjaz2009}, and EP240414a\cite{SunH24}. Light curves of the low-luminosity GRB\,100316D and XRF\,060218 were retrieved from the UK \textit{Swift} Science Data Centre\cite{2009MNRAS.397.1177E}, with K-corrections applied using their time-resolved energy spectra. The shaded region marks the interval during which EP250108a was occulted by the Earth. Triangle symbols indicate upper limits from EP/WXT or EP/FXT.
}
\label{fig:Xray_lc}
\end{figure}

\begin{figure}
\centering
\begin{tabular}{c}
\includegraphics[width=0.8\textwidth]{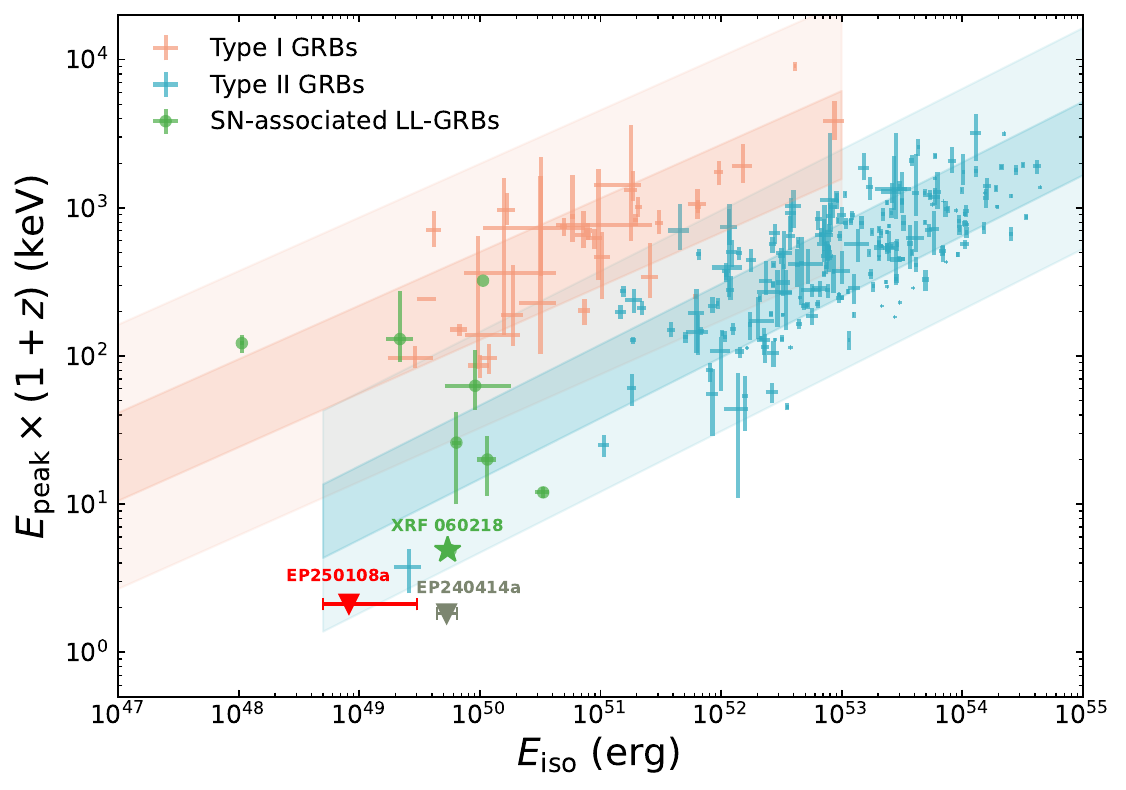}
\end{tabular}
\caption{\noindent\textbf{Location of EP250108a in the $E_{\rm peak}(1+z)$–$E_{\rm iso}$ plane relative to other GRB/FXTs.} Rest-frame spectral peak energy versus isotropic-equivalent energy correlation, known as the ``Amati relation.'' Type I (orange) and Type II (blue) GRBs are classified based on their distinct physical origins: Type I GRBs originate from compact-object mergers, while Type II GRBs result from the core collapse of massive stars. The darker and lighter shaded regions indicate the 1$\sigma$ and 3$\sigma$ scatter around the best-fit relation (see methods), respectively. GRB data are compiled from the literature\cite{Amati08,Zhang09,Nava12,Tsvetkova17,Minaev20,Kienlin20,Zou18}. The SN-associated LL-GRBs listed in Table 1 of Ref.\cite{Rudolph2022} are also plotted with data collected from the literature\cite{Kaneko2007,Ghisellini2006,Zhang09,Starling2011,Cano2017,DElia2018,Chand2020,Fletcher2020,Minaev2020b}. 
For EP250108a and EP240414a, the triangle symbols represent upper limits on $E_{\rm peak}$, and their $E_{\rm iso}$ values are calculated in the 0.5--4\,keV band, which dominates the total energy in the 1--$10^4$\,keV range owing to their soft spectra\cite{SunH24}. The $E_{\rm iso}$ uncertainty for EP250108a reflects both the observed emission and the inferred contribution during the Earth-occultation interval (see Fig.~\ref{fig:Xray_lc}b and Methods). Uncertainties are given at the 1$\sigma$ confidence level.}
\label{fig:amati}
\end{figure}

The prompt X-ray emission of EP250108a displays a smooth pulse (Fig.~\ref{fig:Xray_lc}a). Owing to Earth's occultation, the later observations were interrupted, leaving a gap from $T_0 + 1053$\,s to $T_0 +3937$\,s (Fig.~\ref{fig:Xray_lc}b). Therefore, the duration of the X-ray burst, quantified as $T_{90}$—the time interval during which 90\% of the burst’s total fluence is observed—is $960_{-208}^{+3092}$\,s (Table~\ref{tab:obs_prob}). The time-integrated spectrum of the prompt emission is well described by an absorbed power-law model with a photon index $\alpha = -2.75 \pm 1.1$, placing an upper limit on the peak energy of $E_{\rm peak} \lesssim 1.8\,\mathrm{keV}$ (see Methods). At $z = 0.176$, the absorbed peak isotropic X-ray luminosity in the $0.5$–-$4\,\mathrm{keV}$ band is $\sim 1.8 \times 10^{46}\,\mathrm{erg\,s^{-1}}$, comparable to that of XRF\,060218\cite{Campana+2006}. As shown in Fig.~\ref{fig:Xray_lc}c, the X-ray luminosity of EP250108a exceeds that of XRO\,080109\cite{Modjaz2009} by more than two orders of magnitude, but remains lower than that of classical LL-GRBs --- typically interpreted as resulting from relativistic shock breakouts\cite{Wang_2007,Nakar2012} --- as well as EP240414a, an FXT likely powered by a weak relativistic jet\cite{SunH24}. The total isotropic-equivalent X-ray energy of EP250108a in the 0.5–-4\,keV band, $E_{\rm iso}$, is estimated to lie between $5.0 \times 10^{48}$ and $3.0 \times 10^{49}$\,erg, reflecting the uncertainty in the event’s duration (see Fig.~\ref{fig:Xray_lc}b and Methods). EP250108a lies within the 3$\sigma$ scatter region of the Amati relation\cite{Amati02} for Type II GRBs, as does XRF\,060218, while EP240414a falls outside this range (Fig.~\ref{fig:amati}). EP250108a is a softer and weaker sibling of XRF\,060218, and extends the Amati relation --— commonly applied to LGRBs, XRFs, and LL-GRBs\cite{Amati02,Zhang2009} —-- to the softest and weakest end observed to date (Fig.~\ref{fig:amati}). Its prompt emission is likely powered by a relativistic outflow, but whether it originates from a relativistic shock breakout or a successfully emerged jet remains unsettled.

Follow-up observations in the X-ray and radio bands revealed no detectable emission from EP250108a/SN\,2025kg. X-ray observations with FXT and \textit{Swift}/XRT yielded nondetections, with upper limits that constrain any high-energy afterglow emission (Extended Data Table \ref{tab:FXT_obs}). Assuming a temporal decay similar to that of XRF\,060218, the results suggest a rapidly fading or intrinsically faint X-ray counterpart. Radio observations with ATCA, VLA, and MeerKAT across multiple epochs and frequencies (3--10\,GHz) also revealed no emission (Extended Data Table \ref{table:obs_radio}). The derived radio luminosity limits are comparable to those of LL-GRBs such as XRF\,060218/SN\,2006aj and GRB\,980425/SN\,1998bw, and are significantly below those observed in classical high-luminosity events (Extended Data Figure~\ref{fig:radio_LC}). These deep nondetections effectively rule out a powerful on-axis relativistic jet, pointing instead to an off-axis geometry or a fundamentally different explosion mechanism.

\begin{figure}
\centering
\begin{tabular}{c}
\includegraphics[width=0.8\textwidth]{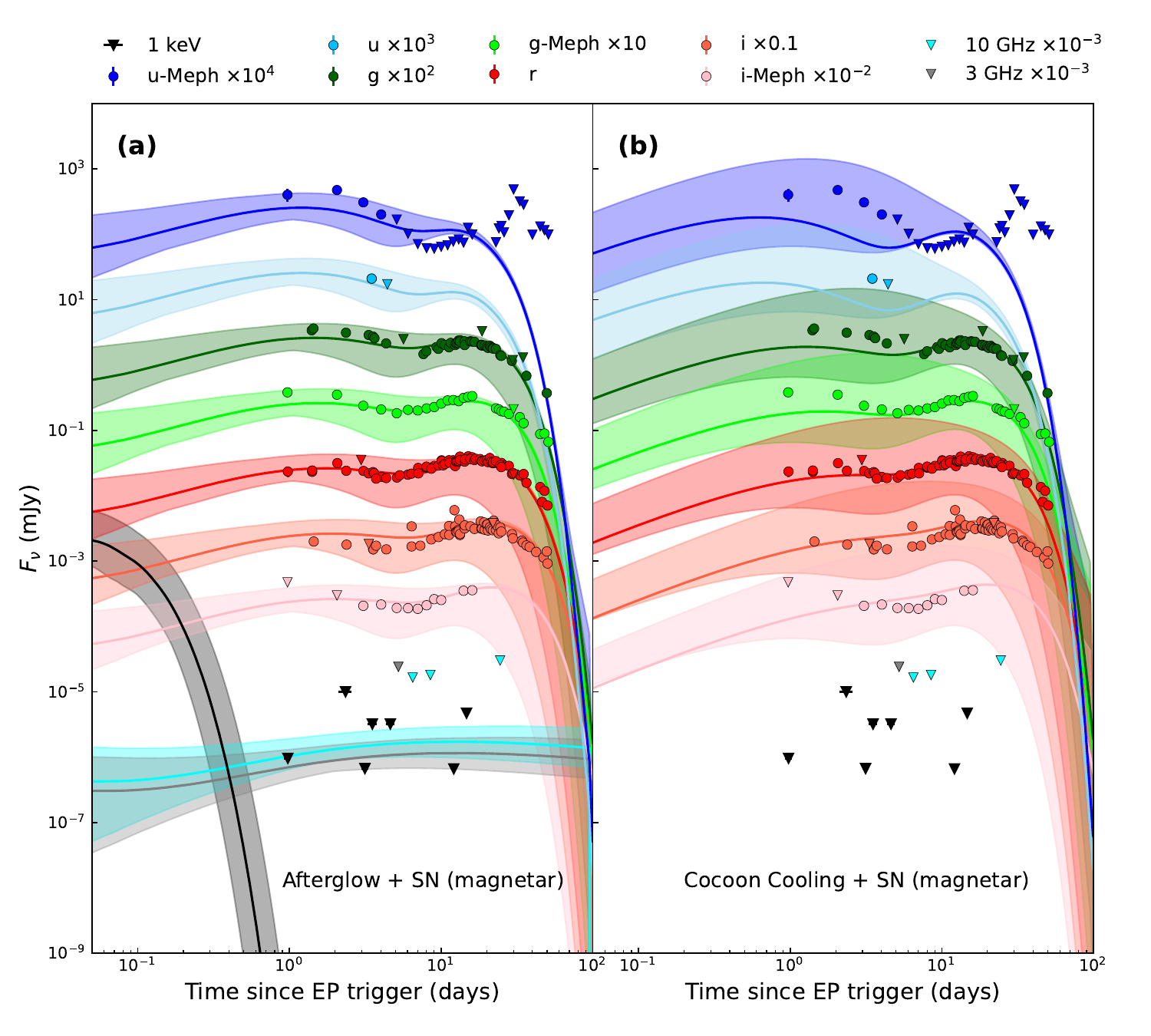}
\end{tabular}
\caption{\noindent\textbf{Multiwavelength light curve of EP250108a and model interpretations.} 
The emission consists of an early-time component, attributed to either (a) jet afterglow or (b) cooling of cocoon material, and a late-time supernova component (see Methods). 
The SN emission is modeled using a magnetar-powered scenario. 
Data points in different frequency bands are shown in different colors, with $1\sigma$ uncertainties indicated. Triangles represent $5\sigma$ upper limits. Shaded regions denote the $1\sigma$ confidence intervals of the model light curves derived from Bayesian posterior distributions, while solid lines indicate the best-fit median models. 
Optical data have been corrected for Galactic extinction, assuming $E(B-V) = 0.02$\,mag and $R_V = 3.1$.
}
\label{fig:slowjet}
\end{figure}

Upon receiving the GCN alert\cite{2025GCN.38861....1L}, we initiated optical follow-up observations. Approximately one day after the EP trigger, the 1.6\,m Mephisto telescope detected the optical counterpart of EP250108a, subsequently named SN\,2025kg\cite{2025GCN.38878....1E}, across multiple wavelengths\cite{2025GCN.38914....1Z}. Following the detection of SN\,2025kg, we coordinated additional follow-up observations (see Methods for details). The follow-up campaign continued until $T_{0} + 51$ days, concluding when the source's elevation became too low for ground-based observations. The photometric evolution of SN\,2025kg is shown in Fig.~\ref{fig:slowjet}, and the full set of photometric measurements obtained during this period is presented in Extended Data Table~\ref{tab:opt_ph}. The multiband light curves exhibit diverse evolution across different filters; however, the overall trend can be roughly divided into two phases: a shallow decay phase before $\sim$5 days, followed by a slow rise. During the initial phase (Phase I; $T_0$ to $\sim T_0 + 5$ days), the light curves exhibit an approximately flat evolution, characterized by a shallow decay slope of $\alpha \approx -0.1$. This is followed by a transition to a steeper decline beginning around $T_0 + 2.5$ days. The light curves exhibit a clear wavelength dependence, with shorter wavelengths decaying more rapidly (see Methods). Beginning around $T_0 + 5$ days (Phase II), SN\,2025kg enters a slow rise phase, reaching peak luminosity at approximately $T_0 + 14.5$ days with an absolute magnitude of $M_g = -19.5$\,mag. Following the peak, the multiband light curves enter a declining phase, exhibiting a temporal evolution similar to that of SNe Ic-BL, in agreement with the spectroscopic classification (see below).

\begin{figure}
\centering
\begin{tabular}{c}
\includegraphics[width=0.8\textwidth]{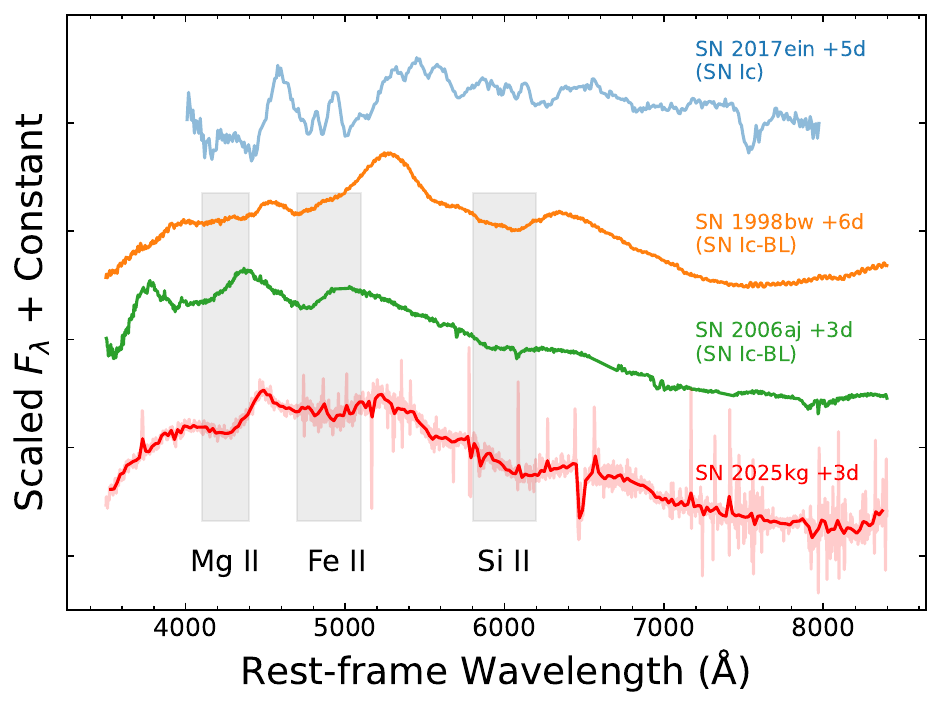}
\end{tabular}
\caption{\noindent\textbf{Comparison of the optical spectrum of SN\,2025kg with spectra of some stripped-envelope SNe.} The Keck~I spectrum of SN\,2025kg, obtained at $T_0\,+\,17.7$ days, is rebinned to a width of 20\,\AA\ (red) for clarity. It is compared with spectra of SNe Ic-BL SN\,1998bw\cite{2001ApJ...555..900P} and SN\,2006aj\cite{Pian2006Nature}, as well as the normal Type Ic SN\,2017ein\cite{2021MNRAS.502.3829T}.
Rest-frame phases, relative to the epoch of peak luminosity, are indicated next to each comparison spectrum. Broad absorption features characteristic of SNe Ic-BL are highlighted in light grey. The trough around 6500\,\AA\ in the SN\,2025kg spectrum is due to telluric absorption lines.
}
\label{fig:sp_com}
\end{figure}

An optical spectrum of SN\,2025kg obtained with ALFOSC on the Nordic Optical Telescope at $T_0\,+\,10$ days closely resembles those of the SNe Ic-BL SN\,1998bw and SN\,2006aj\cite{2025GCN.38984....1X}. Additional spectra were acquired with various telescopes; details are provided in Extended Data Table~\ref{tab:opt_spec} and shown in Extended Data Figure~\ref{fig:opt_sp}. The spectrum having the highest signal-to-noise ratio (SNR), obtained with LRIS on Keck~I at $T_0 + 17.7$ days (near peak brightness), is shown in Fig.~\ref{fig:sp_com}. It closely matches the spectra of SNe Ic-BL, yielding a best-fit $z = 0.17 \pm 0.01$ (see Methods). The spectra of SN\,2025kg exhibit prominent absorption features at 4100–-4400\,\AA, 4800–-5200\,\AA, and 5800–-6000\,\AA, corresponding to \mgii\,$\lambda\,2800$, blended \feii\ lines, and \siii\,$\lambda\,6355$, respectively. These features are characteristic of SNe Ic-BL and support the classification of SN\,2025kg. The absence of hydrogen and helium lines further suggests that the progenitor star was heavily stripped of its outer envelopes. Taken together, the spectroscopic properties indicate that SN\,2025kg originated from the core collapse of a Wolf-Rayet star\cite{Crowther2007}.

A faint host galaxy ($r = 23.2$\, mag) was identified at an angular offset of $0.3''$  from EP250108a/SN\,2025kg\cite{Dey+2019}, corresponding to a projected physical offset of $0.9 \pm 0.1$\,kpc. The absolute magnitude of the host ($M_r = -16.5$) is significantly fainter than those of typical SN Ic-BL host galaxies\cite{Zou+2018,Qin2024}. Spectra revealed strong, narrow emission lines from the host, allowing for a measurement of $z = 0.176 \pm 0.002$. Using the $R_{23}$ strong-line diagnostic\cite{1979MNRAS.189...95P}, we estimate a gas-phase metallicity of $12 + \log(\mathrm{O/H}) = 8.67 \pm 0.13$, consistent with the typical values found for other SNe Ic-BL \cite{Graham2013,Qin2024}. 

The softness and weakness of the prompt emission suggest that EP250108a was powered by a slower and less energetic jet (whether successful or not) compared to typical GRBs. Such a scenario may arise when the jet duration is comparable to the time it takes to traverse the stellar envelope, resulting in a marginally successful breakout. During propagation, the jet deposits a fraction of its energy into the surrounding cocoon material\cite{Bromberg11}, which can subsequently radiate following the breakout of the cocoon shock\cite{Nakar10,Waxman07,Piran19,Gottlieb22}. Detailed modelling shows that the Phase I emission can be interpreted either as the afterglow from a jet with a moderate Lorentz factor of a few tens (Fig.~\ref{fig:slowjet}a) or as shock-cooling emission from the cocoon (Fig.~\ref{fig:slowjet}b; Methods). 
Although the possibility that the jet is viewed off-axis cannot be ruled out, the viewing angle should not deviate from the jet core significantly owing to the absence of the continuous brightening multiwavelength afterglow in Phase I\cite{Lazzati18}.
These characteristics establish EP250108a as a distinct member of the collapsar-origin X-ray transient population.

The emission in Phase II is dominated by the SN Ic-BL. Its light curve --- shaped by energy input from shocks or internal power sources such as radioactive decay or magnetar spindown, diffusing through expanding, optically thick ejecta --- is well described by Arnett-like models\cite{Arnett1982}. Fitting the multiband data yields consistent ejecta mass (\(M_{\rm ej} \approx 2.4\,M_{\odot}\)) and kinetic energy (\(E_{\rm K} \approx 1.2\text{--}1.4 \times 10^{52}\,\mathrm{erg}\)) in both the radioactive decay-powered case (\(M_{\rm Ni} = 0.77\,M_{\odot}\)) and the magnetar-powered case (\(P \approx 14.5\,\mathrm{ms}\), \(B \approx 2.6 \times 10^{14}\,\mathrm{G}\)), each reproducing the observed light curve equally well. However, the unusually high \(M_{\rm Ni}/M_{\rm ej}\) ratio inferred from the radioactive decay-powered scenario still seriously challenges the standard understanding of nucleosyntheses during core-collapse SNe. 
It is alternatively suggested that such SNe associated with FXTs are very likely to be significantly influenced by a central engine. This further hints that the central engine could also play an important role in the powering of SNe following XRFs and even normal GRBs, as previously suggested for SN\,2011kl \cite{Greiner2015} and SN\,2006aj \cite{Zhang2022}. 

EP250108a exhibits remarkable similarities to XRF\,060218, including comparable peak X-ray luminosity, analogous prompt emission duration, and a spectroscopically confirmed association with an SN Ic-BL. The event rate density of EP250108a aligns with that of LL-GRBs possessing luminosities similar to that of XRF\,060218 (Methods). The associated supernova, SN\,2025kg, is a luminous SN Ic-BL, potentially powered by the rotational energy from magnetar spindown. Notably, EP250108a extends the Amati relation to unprecedentedly soft and weak extremes. Modeling of the long-term optical light curves favors interpretations involving either a successful on-axis coasting jet or the shock cooling of a cocoon surrounding a failed jet, both supporting a relativistic outflow origin for this fast X-ray transient.

\clearpage

\clearpage

\clearpage

\clearpage

\clearpage
\begin{table*}
\centering
\begin{threeparttable}
\caption{\textbf{Properties of EP250108a.} Errors represent the 1$\sigma$ uncertainties.}
\label{tab:obs_prob}
\begin{tabular}{lcc}
\toprule
 &  EP250108a\tnote{*} & XRF\,060218\tnote{**} \\
\hline
\textbf{X-Ray} & \\
\hline
Duration$^{\dag}$ $T_{90}$ (s) & $960^{+3092}_{-208} $ &  $2100\pm100$  \\
Photon Index $\alpha$ & $-2.75 \pm 1.1$ &  $\alpha_1\approx-1.0, \alpha_2\approx-2.5$\\
Intrinsic Absorption $N_{\rm int}$ ($\rm cm^{-2}$ ) &  $(4.04 \pm 3.63) \times 10^{21}$ & $(6.3\pm0.3)\times10^{21}$\\
Peak Energy ($\rm keV$) & $< 1.8$ & $4.9\pm0.3$\\
Peak Flux ($\rm erg\,cm^{-2}\,s^{-1}$) & $(1.8_{-0.6}^{+1.8}) \times10^{-10} $  & $\sim1\times10^{-8}$\\

\hline
\textbf{Associated Supernovae} & \\
\hline
Supernova & SN\,2025kg & SN\,2006aj  \\
Type & Ic-BL & Ic-BL \\
Peak Time (day) & $ 14.5$ & 10.4\\
Peak Brightness &  -19.5 $M_g$ & -18.7  $M_V$ \\
\hline
\textbf{Host Galaxy} & \\
\hline
Absolute Brightness &  -16.5 $M_r$ &  -16.15 $M_V$\\
Redshift & $0.176$ & 0.033\\
Projected Offset ($\rm kpc$) &   \(0.9 \pm 0.1\)\, & $\sim$2.2\\

\hline
\textbf{Derived Properties} & \\
\hline
X-ray Peak Luminosity ($\rm erg\,s^{-1}$) & $(1.8_{-0.7}^{+2.6})\times10^{46}$ & $\sim1.5\times10^{46}$\\ 
Isotropic Energy$^{\dag\dag}$ ($\rm erg$) & $ 8.2_{-3.2}^{+21.8} \times10^{48} $ & $(5.3\pm0.03)\times10^{49}$\\ 
\bottomrule
\end{tabular}
\begin{tablenotes}
\footnotesize
\item[*] The J2000 source position is R.A. = $03^{\rm hr}42^{\rm m}28.40^{\rm s} \pm 0.05^s$, Dec. = $-22^\circ30'21.2'' \pm 0.1''$, inferred from the NOT observation. X-ray properties of EP250108a are in the observed 0.5--4\,keV band. The derived properties of EP250108a are in the rest-frame 0.5-4 keV band, which dominates that of 1--10$^4$\,keV in the rest frame given the soft photon index.
\item[**] Properties of XRF\,060218/SN2006aj are collected from the literature. Duration, peak energy, peak flux: Campana et al. (2006)\cite{Campana+2006}; photon index: Toma et al. (2007)\cite{Toma+2007}; intrinsic absorption: Liang et al. (2006)\cite{Liang+2006}; supernova peak time, peak brightness, and host brightness: Sollerman et al. (2006)\cite{Sollerman+2006}; peak luminosity is derived from the peak flux, and its energy is in the rest-frame range 0.5--4\,keV, to compare with the X-ray luminosity of EP250108a; isotropic energy: Amati et al. (2008)\cite{Amati08}
\item[\dag] The central value represents the observed $T_{90}$, and the upper limit considers both the error of observed duration and the Earth-occultation time.
\item[\dag\dag] The central value represents the observed isotropic energy, and the upper limit considers the model error and the inferred maximum isotropic energy during the Earth-occultation period; see Methods.

\end{tablenotes}
\end{threeparttable}
\end{table*}

\clearpage

\section*{Methods}\label{sec:methods}
\subsection{X-ray Observations and data reduction.\\} \label{sec:obs}

The Einstein Probe (EP) mission \cite{Yuan2025}, led by the Chinese Academy of Sciences (CAS) with the European Space Agency (ESA) and the Max Planck Institute for Extraterrestrial Physics (MPE), has been in orbit for more than one year. The Wide-field X-ray Telescope \cite{EPhandbook} (WXT; 0.5--4\,keV) and the Follow-up X-ray Telescope \cite{FXT} (FXT; 0.3--10\,keV) are the two main instruments mounted on the EP. The WXT's novel lobster-eye micropore optics (MPO) enable a large instantaneous field of view (FOV) of 3600 square degrees and a sensitivity of $\sim\,2.6 \times 10^{-11}\, \rm erg\,cm^{-2}\,s$ in the 0.5--4\,keV band for an exposure time of 1\,ks. FXT comprises two coaligned modules (FXT-A and FXT-B), each containing 54 nested Wolter-I paraboloid-hyperboloid mirror shells.

EP250108a was detected by the WXT and reported by the offline archive search pipeline; see the WXT image in Extended Data Figure~\ref{fig:X_ray_detail}a. The redefined start time of the event is 2025-01-08 UTC 12:47:35.72 by using the Bayesian block method \cite{Bayesianblock} to rebin the photons with 50\,ms temporal resolution; see Fig.~\ref{fig:Xray_lc}a. The processing and calibration of WXT photon events are handled using specialized data-reduction software and the calibration database (CALDB) (Y. Liu et al., in prep.). The calibration database is generated based on the results of the on-ground calibration experiments (H.-Q. Cheng et al., in prep.). The position of each photon was converted to celestial coordinates (J2000). The energy of each event is calculated using the bias and gain values stored in the calibration database. After flagging bad and flaring pixels, only single-, double-, triple-, and quadruple-pixel events without anomalous flags were selected to generate the cleaned event file. Source photons were extracted from a circular region with a radius of $9'$, while background photons were extracted from an annular region with inner and outer radii of $18'$ and $36'$, respectively. As the average net count rate of WXT during the total prompt emission phase is  $\sim 0.049~\mathrm{counts~s^{-1}}$, we grouped the WXT data with a minimum of two counts per bin for spectral analysis. The integrated spectrum during the $T_{90}$ interval was modeled using \texttt{Xspec v12.14.0h}\cite{Xspec}, adopting the model $ztbabs\times tbabs\times powerlaw$. In this model, \texttt{ztbabs} accounts for intrinsic absorption ($N_{\mathrm{int}}$) and \texttt{tbabs} indicates the Galactic absorption. The T$\mathrm{\ddot{u}}$bingen–Boulder interstellar matter (ISM) absorption model\cite{TbAbs_model} is used for the two hydrogen photoelectric absorptions above. The \texttt{powerlaw} represents a power-law spectrum of the form $N(E) = K\times(E/1\,\mathrm{keV})^{\alpha}$, where $K$ is the normalization and $\alpha$ is the photon index. The Galactic  hydrogen column density and the redshift of EP250108a are fixed to $ 1.60 \times 10^{20}\,\rm{cm}^{-2}$ \cite{NH_method} and 0.176, respectively. A time-averaged $N_{\mathrm{int}}=(4.04 \pm 3.63) \times 10^{21} \,\rm{cm}^{-2}$ and a photon index of $\alpha = -2.75 \pm 1.1$ were obtained, with an acceptable fit statistic of CSTAT/(d.o.f.) $\approx$ 18.7/22; see Extended Data Figure.~\ref{fig:X_ray_detail}b. Furthermore, we consider an absorbed blackbody model $ztbabs \times tbabs \times blackbody$. The intrinsic absorption cannot be well constrained, with a 90\% confidence level upper limit of $N_{\mathrm{int}}<4.7 \times 10^{21} \,\rm{cm}^{-2}$. The derived temperature is $0.34 \pm 0.07$\,keV with a fit statistic of CSTAT/(d.o.f.) $\approx$ 18.0/22, showing no significant improvement compared to the power-law model.

The soft spectrum, modeled with a single power law, suggests that the peak energy $E_{\rm peak}$ lies near or below 0.5\,keV, the lower bound of WXT's energy range. To better constrain $E_{\rm peak}$, we performed spectral fitting with an absorbed broken power-law model. The low-energy photon index was fixed at $-1$, a typical value for GRBs, while the high-energy index was derived as $-4.2^{+1.6}_{-5.1}$. The intrinsic absorption was constrained to be $N{\mathrm{int}} < 3.9 \times 10^{21}\,\mathrm{cm}^{-2}$ at the 90\% confidence level. The peak energy cannot be well constrained. An upper limit at the 90\% confidence level of $E_{\rm peak} < 1.8$\,keV is obtained from the absorbed broken power-law model, with a statistic of $\mathrm{CSTAT}/\mathrm{d.o.f.} \approx$ 16.7/21. This result suggests that $E_{\rm peak}$ likely lies below 1.8\,keV. All quoted uncertainties from spectral fitting correspond to the 1$\sigma$ confidence level unless otherwise specified.

Follow-up X-ray observations were conducted with the FXT and the \textit{Swift} X-Ray Telescope (XRT) from $T_0 + 22$\,hr to $T_0 + 15$\,days. No X-ray emission was detected in any of the epochs, and the corresponding upper limits are provided in Extended Data Table~\ref{tab:FXT_obs}. Since no X-ray emission was detected beyond the WXT detection of EP250108a, we assumed a power-law decay index of $-5.5$ for the late-time light curve, consistent with that observed in XRF\,060218\cite{Margutti2015}, and potentially applicable to EP250108a. Intrinsic absorption was not included in the calculation owing to the lack of constraints from the available data.

\subsection{X-ray Analysis.\\} \label{sec:x-ray-analysis}
To compare the energetics and spectral properties of EP250108a with those of other GRBs, particular attention is given to the empirical correlation between the rest-frame peak energy, $E_{p,z} = E_{\mathrm{peak}} \times (1+z)$, and the isotropic-equivalent gamma-ray energy release, $E_{\gamma,\mathrm{iso}}$, known as the Amati relation. This correlation is described by the expression $\log E_{p,z} = b + k \log E_{\gamma,\mathrm{iso}}$. The relation was fitted to samples of both Type I and Type II GRBs using the Python package \texttt{emcee} \cite{Foreman-Mackey2013}. The best-fitting parameters and their $1\sigma$ uncertainties are $k = 0.36^{+0.05}_{-0.05}$ and $b = -15.70^{+2.50}_{-2.76}$ for Type I GRBs, and $k = 0.41^{+0.02}_{-0.02}$ and $b = -19.08^{+1.32}_{-1.32}$ for Type II GRBs. The $1\sigma$ scatter was estimated from the standard deviation of the residuals between observed and predicted $E_{\gamma,\mathrm{iso}}$ values. Luminosity distances were calculated assuming the cosmological parameters reported by the Planck Collaboration\cite{Planck2020}. 
EP250108a lies within the $3\sigma$ region of the best-fit Amati relation while its upper limit on $E_{\gamma,\mathrm{iso}}$ exceeds the $3\sigma$ scatter range.

We investigate the similarity between EP250108a and XRF\,060218 in the soft X-ray band. To assess the detectability of XRF\,060218 by EP/WXT, we place it at $z = 0.176$ and compare its observed luminosity, X-ray spectrum, light curve, and absorption properties to those of EP250108a. The original unabsorbed flux light curve of XRF\,060218 is adopted from the Burst Analyser, using data points with an SNR of at least 5 \cite{Evans2010}. The photon index in each \textit{Swift}/XRT time bin is obtained directly from the Burst Analyser, while for \textit{Swift}/BAT time bins it is fixed at $-1.4$, consistent with the value measured during the overlapping interval between BAT and XRT observations. We simulate the source and background spectra using the \texttt{fakeit} command in \texttt{XSPEC}, adopting the best-fitting continuum model of XRF\,060218 for each time bin. The same exposure time, redistribution matrix, and auxiliary file for each of the time bins are used in these simulations to mimic the observed spectra, and then simulated spectra are generated following a Poisson distribution. The simulated EP/WXT light curve of an XRF\,060218-like event at $z = 0.176$ is shown in Fig.~\ref{fig:Xray_lc}b, rebinned with an SNR of 3. 

As illustrated in Fig.~\ref{fig:Xray_lc}b, we estimate the X-ray flux of EP250108a during the Earth-occultation period. Assuming that the emission persisted throughout the entire occultation, a reasonable maximum flux scenario is that the source maintained its average observed flux and then decayed with the same temporal slope as XRF\,060218, down to the observed upper limit. Under this assumption, the isotropic energy emitted before and during the occultation period is taken as the upper limit of the total $E_{\rm iso}$ of EP250108a, estimated to be $3\times10^{49}$\,erg.

\noindent\textbf{Gamma rays.}
The \textit{Fermi} Gamma-ray Burst Monitor \cite{GBM} (GBM) was not occulted by Earth since around $\sim \,T_0-610$\,s for EP250108a. A blind search for a GRB-like signal was run in the time interval ${T_0}^{+1000}_{-500}$\,s from timescales of 0.256 to 16.384\,s, and no significant signal was found related to EP250108a, consistent with the targeted search result \cite{2025GCN.39146....1R}. With the response files generated by \texttt{GBM Response Generator} (\url{https://fermi.gsfc.nasa.gov/ssc/data/analysis/gbm/DOCUMENTATION.html}) for the location of EP250108a around $T_{0}-600$\,s to $T_{0}+1000$\,s, we can derive the average 3$\sigma$ upper limit in the 10--1000\,keV band with a power-law index $-2.75$ to be $\sim1.7 \times 10^{-8}\, \rm erg\,cm^{-2}\,s^{-1}$ for a 8.192\,s timescale. 

\subsection{Optical photometry.\\} \label{sec:optical_ph}
Optical photometric observations were conducted using multiple facilities. The Mephisto telescope continuously monitored the counterpart from its initial detection until it became unobservable. Through the Global Supernova Project\cite{Howell2017}, $griz$-band images obtained using the global network of 1.0\,m telescopes operated by the Las Cumbres Observatory (LCO)\cite{Brown2013} provided continuous coverage throughout the observing campaign, with contributions from sites including the Siding Spring Observatory, South African Astronomical Observatory, Cerro Tololo Inter-American Observatory, McDonald Observatory, and Haleakala Observatory. Additional photometric data were obtained from the Nordic Optical Telescope (NOT), the Lijiang 2.4\,m Telescope (LJT), the Liverpool Telescope (LT), the SAO RAS Zeiss-1000 Optical Telescope (Zeiss-1000), the Southern Astrophysical Research (SOAR) Telescope, the 1.0\,m Lesedi Telescope at the South African Astronomical Observatory (SAAO), the 10.4\,m telescope Gran Telescopio Canarias (GTC), the 2.2\,m telescope on the Centro Astron\'{o}mico Hispano en Andaluc\'{i}a (CAHA), the  10\,m Keck~I telescope, and the 60\,cm Rapid Eye Mount (REM) telescope in Chile. 
Along the line of sight of EP250108a/SN\,2025kg, the Galactic extinction is $E_{B-V}=0.015$\,mag \cite{Schlafly2011}. 

Mephisto is a 1.6\,m wide-field, multichannel photometric survey telescope developed and operated by the South-Western Institute for Astronomy Research at Yunnan University. It is located at the Lijiang Observatory (IAU code: 044) of the Yunnan Astronomical Observatories, Chinese Academy of Sciences. Equipped with three CCD cameras, Mephisto enables simultaneous imaging in either the $ugi$ or $vrz$ bands during each observation\cite{Yang24, Chen2024ApJ}. 
Simultaneous multiband photometry of EP250108a/SN\,2025kg with Mephisto began about one day after the discovery and continued until $T_0 + 51$\,days. The transient was detected in the $uv$ bands before $T_0 + 4.03$\,days and remained visible in the $gr$ bands throughout the campaign. It was visible in the $i$ band from $T_0 + 3.07$ to $T_0 + 16.05$\,days, with no further data thereafter. No detection was made in the $z$ band during the entire monitoring period.

The raw frames were reprocessed using a dedicated pipeline developed for Mephisto, including bias and dark subtraction, flat-field correction, and cosmic-ray removal. Source detection was carried out with \texttt{SExtractor}\cite{Bertin1996}, followed by astrometric calibration using reference stars from the Gaia DR3 catalog\cite{GaiaDR3}. The resulting astrometric uncertainties are estimated to be better than 0.1 pixels.

Multiple exposures in each band from the same night were coadded using \texttt{SWarp}\cite{Bertin2010ascl} to improve the SNR. A point-spread function (PSF) model was constructed for each stacked image using \texttt{SExtractor} and \texttt{PSFex}\cite{Bertin2011}, and PSF photometry was performed on the coadded images. For sources that were not clearly detected, the 5$\sigma$ limiting magnitudes were estimated from the coadded images. The zero-point values were derived from the reference images, with the photometric calibration performed using the synthetic-photometry method based on the recalibrated Gaia BP/RP (XP) spectra \cite{Huang2024ApJS, Xiao2023ApJS}. The 1$\sigma$ uncertainties associated with the photometric calibration are estimated to be better than 0.03\,mag in the $u$ band, 0.01\,mag in the $v$ band, and 0.005\,mag in the $griz$ bands.

Two instruments mounted on the 1\,m Zeiss-1000 telescope were used to observe the optical counterpart of EP250108a: a CCD photometer\cite{2020AstBu..75..486K} and MAGIC\cite{2020AzAJ...15a...7K}. Observations were carried out in the {\it B, V, Rc}, and {\it Ic} filters. On nights with stable weather and good seeing, a series of 300\,s exposures was taken in each filter; otherwise, only the {\it Rc} band was used. The data were reduced using standard ESO-MIDAS procedures, including bias subtraction, flat-field correction, and cosmic-ray removal. To eliminate fringing in the {\it Rc} and {\it Ic} images, small dithers were applied between exposures. Median-combined fringe templates were created from each dithered set and subtracted from the science frames. Images from each night were aligned to a common coordinate system and stacked to improve the SNR.

Observations of SN\,2025kg were obtained using the Mookodi spectrograph/imager\cite{2024JATIS..10b5005E} on the SAAO 1.0\,m Lesedi telescope using 120\,s exposures in \textit{Sloan} $g'$, $r'$, and $i'$ filters on four consecutive nights (19--22 Jan. 2025). The observations were taken under the SAAO's Intelligent Observatory transient follow-up programme\cite{2024SPIE13098E..0YP}. 

The raw frames obtained with LCO were automatically processed using the \textit{BANZAI} pipeline (\url{https://github.com/LCOGT/banzai}). For NOT, LJT, LT, SOAR, and REM, raw images were reduced with standard \textit{IRAF} tasks\cite{Tody+1986}, including bias subtraction, flat-field correction, and cosmic-ray removal. Astrometric calibration was performed using \textit{astrometry.net}\cite{Lang+2010}. Aperture photometry for stacked images from LCO, NOT, LJT, LT, Zeiss-1000, SOAR, and REM was carried out with \texttt{SExtractor}\cite{Bertin1996}. The {\em Sloan} $g'r'i'z'$-band photometry was calibrated against nearby reference stars from the Pan-STARRS DR2 catalog in the AB magnitude system\cite{2020ApJS..251....7F}. For the \textit{Johnson-Cousins} filters, the \textit{BVRcIc}-band photometry in the Vega system was calibrated using magnitudes converted from the {\em Sloan} system (\url{https://www.sdss.org/dr12/algorithms/sdssUBVRITransform/#Lupton}). 

The final optical photometric results are summarized in Extended Data Table~\ref{tab:opt_ph}, and the light curves are shown in Fig.~\ref{fig:slowjet}.

To estimate the pseudobolometric light curve of SN\,2025kg, we reconstructed the spectrum obtained with the Keck~I telescope on 26 January. Inspection of the spectral sequence suggests that the spectral shape does not evolve significantly over time; thus, this spectrum was adopted as a representative SED for SN\,2025kg. 

Multiband photometry was used to calibrate the flux scale of the spectrum at the corresponding epoch. Galactic extinction correction was applied\cite{CSFD+2023}, and the dereddened spectrum was then integrated to yield a luminosity of $L = 1.5 \times 10^{43}\,\rm erg\,s^{-1}$. By comparing this result with the $i$-band apparent magnitude at that time, a correction magnitude of $\Delta m_{i} = 39.4$ was derived. Applying this correction to epochs later than 5 days (Phase II) yielded the pseudobolometric light curve shown in Extended Data Figure~\ref{fig:bolometric}. The pseudobolometric light curve of SN\,2025kg is brighter than those of SN\,1998bw\cite{1998Natur.395..670G} and SN\,2006aj\cite{2006Natur.442.1018M}, placing it among the most luminous SNe Ic-BL observed to date. However, it remains fainter than the ultra-luminous GRB-associated SN\,2011kl\cite{2015Natur.523..189G} and the hydrogen-poor superluminous supernova PTF11rks\cite{2013ApJ...770..128I}.

\noindent\textbf{Radio observations.}

Radio follow-up observations of EP250108a/SN\,2025kg were conducted with MeerKAT, Karl G. Jansky Very Large Array (VLA), and Australia Telescope Compact Array (ATCA). MeerKAT observed the source at 3.0\,GHz for 1\,hr on January 13, 2025\cite{2025GCN.38958....1C}. No emission was detected at the target position, resulting in a $3\sigma$ upper limit of 24\,$\mu$Jy\,beam$^{-1}$ (with root-mean-square noise of 8\,$\mu$Jy\,beam$^{-1}$). 
The VLA observations were made on January 15, 2025 (6.5 days post-discovery) at a mean frequency of 10\,GHz\cite{2025GCN.38970....1S}. No radio emission was detected at or near the position of EP250108a/AT2025kg, with a $3\sigma$ upper limit of 16.5\,$\mu$Jy. At a source redshift of $z=0.176$, this corresponds to a 10\,GHz luminosity limit of $<2\times10^{28}$\,erg\,s$^{-1}$\,Hz$^{-1}$. The ATCA observations were carried out on 16 January and 1 February, 2025, at frequencies of 5.5\,GHz and 9.0\,GHz, yielding $3\sigma$ upper limits of 24\,$\mu$Jy and 18\,$\mu$Jy, respectively. When combining data from both spectral windows, a deeper upper limit of 15\,$\mu$Jy ($3\sigma$) was obtained at a central frequency of 7.25\,GHz.

A comprehensive radio monitoring campaign spanning multiple facilities and frequencies revealed no detectable emission from EP250108a/SN\,2025kg. The radio upper limits of EP250108a/SN\,2025kg are comparable to those of XRF\,060218/SN\,2006aj and GRB\,980425/SN\,1998bw, and are significantly lower than those observed in high-luminosity GRBs such as GRB\,030329, GRB\,111209A, and GRB\,130427A (Extended Data Figure~\ref{fig:radio_LC}). 
These stringent upper limits are opposite to the presence of the powerful on-axis ultrarelativistic jet in typical GRBs and the dense surrounding medium. 
The radio upper limits of EP250108a/SN\,2025kg are presented in Extended Data Table \ref{table:obs_radio} and shown in Extended Data Figure~\ref{fig:radio_LC}. 

\subsection{Optical spectroscopic observations.\\ }\label{sec:optical_sp}
A total of 12 optical spectra were obtained with various telescopes worldwide between $T_{0} + \sim10$ days and $T_{0} + \sim50$ days. Details of the observations are provided in Extended Data Table~\ref{tab:opt_spec}, and the spectra are shown in Extended Data Fig.~\ref{fig:opt_sp}.

The first spectrum was obtained on 18.8 January ($T_{0} + 10.4$\,days) using ALFOSC mounted on the NOT. On the same night, an additional spectrum was taken with OSIRIS+ on the GTC (Project PI David Sánchez Aguado, ID GTCMULTIPLE4C-24B). Spectra were also acquired with the GHTS mounted on SOAR on 22 and 24 January, and with the Kast double-beam spectrograph on the 3\,m Shane telescope at Lick Observatory on 30 January. Two further observations were conducted with the NOT on 1.8 February (partially interrupted by bad weather) and 2.9 February.
All of these spectra were reduced using standard \textit{IRAF} procedures, including bias subtraction, flat-field correction, spectral extraction, wavelength calibration, and flux calibration.  

The Southern African Large Telescope (SALT\cite{Buckley+2006}) obtained spectra using the Robert Stobie Spectrograph (RSS\cite{2003SPIE.4841.1463B}) with the PG0700 grating, covering the wavelength range 3590--7480\,\AA\ at a mean resolution of 7.6\,\AA. Two consecutive 1200\,s exposures were taken on each of two nights, 19 and 22 January. The data were reduced with the PyRAF-based PySALT package (\url{https://astronomers.salt.ac.za/software/}\cite{2010SPIE.7737E..25C}), which performs gain and cross-talk corrections, as well as bias subtraction. Spectral extraction was carried out using standard \texttt{IRAF} tasks, including wavelength calibration, background subtraction, and one-dimensional (1D) extraction. Owing to SALT's design, absolute flux calibration is not feasible, as the entrance pupil moves relative to the primary mirror array during observations, causing the telescope's effective collecting area to vary\cite{Buckley+2006}. However, by observing spectrophotometric standards during twilight, we performed relative flux calibration, which provides an approximate recovery of the spectral shape and relative line strengths.

On 23.1 January ($T_{0} + 14.5$\,days), one spectrum was obtained with the IMACS short camera mounted on the 6.5\,m Magellan Telescope (Project PI Brenna Mockler). The IMACS spectrum was reduced using standard \texttt{IRAF} routines, including bias subtraction, flat-field correction, aperture extraction, and wavelength calibration based on a HeNeAr comparison lamp exposure taken immediately after the target observation. Flux calibration and telluric correction were performed using spectrophotometric and telluric standard stars observed on the same night. Some spurious features remain, however, such as residuals from the strong O~I sky emission at 5577\,\AA\ and the telluric A-band near 7600\,\AA.

On 26 January and 27 February, we obtained two long-slit spectroscopy of SN\,2025kg with LRIS mounted on the Keck~I 10\,m telescope\cite{Oke+1995}. The data were reduced with the pipeline \textsc{LPipe}. 

SN\,2025kg is classified as a Type Ic-BL SN at $z=0.17\pm0.01$ using template matching with \textsc{GELATO}\cite{2008A&A...488..383H} (gelato.tng.iac.es) and \textsc{Superfit}, consistent with the redshift of its host galaxy (see below). The spectra exhibit three broad absorption features: \mgii\ ($\sim$4000-–4400\,\AA), a blend of \feii\ lines ($\sim$4500-–5100\,\AA), and an absorption feature around \siii\ ($\sim$5800-–6100\,\AA), as shown in Fig.~\ref{fig:sp_com}. Fitting the \siii\, absorption feature with a Gaussian profile yields an expansion velocity of $v_{\rm Si\,II} \approx 12,000\,\rm km\,s^{-1}$ near peak brightness, consistent with that of other SNe Ic-BL\cite{Finneran+2024}.

\subsection{Host galaxy.\\}
A known source at J2000 coordinates R.A. = $03^{\rm hr}42^{\rm m}28.38^{\rm s}$, Dec. = $-22^\circ30'21.3''$, close to the target EP250108a/SN\,2025kg, was found in the DESI Legacy Imaging Surveys at an offset of $0.3''$  and brightness 23.2\ mag in the Sloan $r$ band\cite{Dey+2019}, as shown in Extended Data Figure~\ref{fig:host}a. Utilizing the cosmological parameters from the Planck Collaboration \cite{Planck2020}, the corresponding luminosity distance and the angular diameter distance are determined to be 870\,Mpc and 629\,Mpc at $z = 0.176$, respectively. Therefore, the projected offset between EP250108a/SN\,2025kg and the center of the host is $0.9\pm0.1$\,kpc, and the corresponding \textit{r}-band absolute magnitude is $-$16.5, significantly fainter than the host galaxies of typical Type Ic-BL SNe\cite{Zou+2018,Qin2024}. 

In the spectrum taken on 26 February with the Keck~I telescope, the SN continuum had significantly faded, revealing prominent narrow emission lines from the host galaxy. These include [O\,II]~$\lambda3727$, H$\beta$~$\lambda4861$, [O\,III]~$\lambda4959$, [O\,III]~$\lambda5007$, and H$\alpha$~$\lambda6563$ \ref{fig:host}. We fitted the emission lines using a Gaussian model and the SCIPY\cite{2020NatMe..17..261V} package to determine the redshift. 
The redshift and its uncertainty were derived from the mean and standard deviation of the fitted Gaussian, resulting in $z = 0.176~\pm~0.002$. We corrected for Galactic extinction adopting $E_{B-V}=0.015$\,mag \cite{Schlegel98,Schlafly2011} and assuming the standard Galactic reddening law with $R_V = 3.1$ \cite{1989ApJ...345..245C}.

The metallicity of the host galaxy was estimated using the empirical $R_{23}$ strong-line diagnostic \cite{1979MNRAS.189...95P}, defined as
\[
R_{23} = \frac{[\mathrm{O\,II}]\,\lambda3727 + [\mathrm{O\,III}]\,\lambda4959 + [\mathrm{O\,III}]\,\lambda5007}{\mathrm{H}\beta}\, .
\]
To measure the line fluxes, we subtracted the local continuum by taking the median flux from adjacent regions ($\pm$20--40\,\AA) on both sides of each line and integrated the continuum-subtracted flux. Flux uncertainties were derived by propagating the per-pixel errors across the integration range. This yielded an observed $R_{23} = 3.6 \pm 0.6$.

We then applied the empirical calibration of Ref. \cite{2017MNRAS.465.1384C}, which is based on electron-temperature metallicities derived from a large SDSS galaxy sample. The relationship between the observed $R_{23}$ and the oxygen abundance $x = 12 + \log(\mathrm{O/H})$ is modeled by a third-order polynomial,
\[
\log R_{23} = \sum_3 c_n (x - 8.69)^n\, ,
\]
where 8.69 is the adopted solar value of $12 + \log(\mathrm{O/H})$ \cite{2001ApJ...556L..63A}, and the best-fit coefficients from Ref. \cite{2017MNRAS.465.1384C} are $c_0 = 0.527$, $c_1 = -1.569$, $c_2 = -1.652$, and $c_3 = -0.421$.  
Given $R_{23} = 3.6 \pm 0.6$, we solved this equation numerically within the upper branch of the $R_{23}$ vs. O/H relation. This yielded an estimated gas-phase metallicity of $12 + \log(\mathrm{O/H}) = 8.67 \pm 0.13$, close to the solar value. This metallicity is higher than that of EP240414a/SN\,2024gsa\cite{SunH24}, and is consistent with the typical values found for other SNe Ic-BL \cite{Graham2013,Qin2024}.

\subsection{Theoretical Modelling.\\}  

\noindent \textbf{The on-axis slow jet.}
The absence of prompt gamma-ray emission, combined with the proximity of EP250108a, suggests that the jet's initial bulk Lorentz factor is significantly lower than the typical values of several hundred.
Moreover, the shallow decay phase observed in the early-time optical light curves within the first four days post-burst is consistent with the temporal behavior expected from a slowly coasting jet interacting with the surrounding medium. For a jet with an initial Lorentz factor $\Gamma_0$, an isotropic-equivalent kinetic energy $E_{\mathrm{K,iso}}$, and a half-opening angle $\theta_{\rm j}$, the coasting timescale in a stellar wind environment is given by Ref. \cite{Blandford76}
\begin{equation}
t_{\rm co} \approx 2 \times 10^5 \left(\frac{E_{\mathrm{K,iso}}}{10^{52}~\mathrm{erg}} \right) \left(\frac{A_{*}}{10^{-3}} \right)^{-1} \left(\frac{\Gamma_0}{20}\right)^{-4}~\mathrm{s}\, , 
\end{equation}
where $A_*$ is the dimensionless wind parameter.
In the standard afterglow model\cite{Sari99}, electrons accelerated by the external shock follow a power-law energy distribution \( dN / d\gamma \propto \gamma^{-p} \). Their synchrotron emission can be characterized by introducing microphysical parameters, including the fraction of electrons accelerated (\( f_{\rm e} \)), and the equipartition fractions for electrons and magnetic fields (\( \epsilon_{\rm e} \), \( \epsilon_B \)).
During this phase, the observed synchrotron peak flux density in the optical bands (\( F_{\nu,\rm peak}\)) places an upper limit on \( \Gamma_0 \):
\begin{equation}
\Gamma_0 \le 20 \left( \frac{F_{\nu,\rm peak}}{0.03~\mathrm{mJy}} \right) 
\left( \frac{f_{\rm e}}{0.1} \right)^{-2/3} 
\left( \frac{A_{*}}{10^{-2}} \right)^{-1} 
\left( \frac{\epsilon_{B}}{10^{-3}} \right)^{-1/3}.
\end{equation}
By applying a Bayesian fitting procedure to the early afterglow light curves, we obtain a set of physically reasonable parameters that provide a good fit to the data (Figure~\ref{fig:slowjet}(a)):
$E_{\mathrm{K,iso}} \approx  6 \times 10^{52}$~erg, $\Gamma_0 \approx 20$, $\theta_{\rm j} \simeq 0.07$~rad, $f_{\rm e} \approx 0.1$, $\epsilon_{\rm e} \approx 0.6$, $\epsilon_{B} \approx 0.002$, $p \approx 3$, and $A_{*} \approx 0.007$.

\noindent \textbf{The shock-cooling scenario.} The early-time optical emission may also originate from the shock cooling of the cocoon\cite{Nakar10,Piran19,Gottlieb22}. 
In collapsar-origin events, the propagation of the jet through the stellar envelope results in the formation of a hot cocoon surrounding the jet itself\cite{Bromberg11}. 
This cocoon material, moving at a speed of \( \sim 0.1\,c \), releases its thermal energy (\( E_{\rm bo} \)) once the shock breaks out of the stellar surface.

The duration of the shock-breakout emission from the cocoon can be estimated as Ref. \cite{Waxman07}
\begin{equation}
t_{\rm bo} \approx 1200\,T_{\rm obs,0.1\,\mathrm{keV}}^{-4/7}~E_{\rm bo,50}^{3/7} 
\left( \frac{\kappa}{\tau} \right)^{2/7}\,\mathrm{s}\, ,
\end{equation}
where \( T_{\rm obs} \) is the observed temperature of the breakout emission, 
\( \kappa \approx 0.2~\mathrm{g~cm^{-2}} \) is the Thomson opacity for fully ionized helium, and \( \tau \simeq 1 \) is the optical depth at breakout.
After the breakout, the ejected cocoon material —-- with deposited energy $E_{\rm ej}$ and mass $M_{\rm ej}$ —-- may contribute to optical-band emission during its cooling phase.  The evolution of the photospheric radius and temperature can be approximated as
\begin{align}
r_{\rm ph} &\approx 4.1 \times 10^{14}\, f_{\rho}^{0.26} \left( \frac{\tau}{\kappa} \right)^{-0.12} E_{\rm ej,51}^{0.38} M_{\rm ej, M_{\odot}}^{-0.27} t_{\rm day}^{0.77}\,\text{cm}\, , \\
T_{\rm ph} &\approx 2.6 \times 10^4\, f_{\rho}^{-0.02} E_{\rm ej,51}^{0.02} M_{\rm ej, M_{\odot}}^{-0.03} R_{*,12}^{1/4} t_{\rm day}^{-0.5}\,\text{K}\, ,
\end{align}
where $R_*$ is the progenitor’s stellar radius and $f_{\rho}$ is a dimensionless parameter characterizing the density profile near the stellar surface.
By applying a Bayesian fitting procedure, we derive a set of parameters --- $f_{\rho} \approx 1$, $\tau \approx 1$, $E_{\rm ej} \approx 6 \times 10^{51}$~erg, $M_{\rm ej} \approx 0.1\,M_{\odot}$, and $R_{*} \approx 3 \times 10^{10}$\,cm --- under which the early-time optical light curves can be interpreted as originating from the cooling emission of the cocoon (Extended Data Figure~\ref{fig:shock_cooling}).

\noindent\textbf{Energy source of supernova component.\\} 

The spectra and photometry of SN\,2025kg suggest that it is an SN Ic-BL. The evolution of its light curve is primarily governed by the deposition of energy, which may originate from an outward-propagating shock or internal heating sources such as radioactive decay or a central engine. This energy subsequently diffuses radiatively through the optically thick, expanding ejecta, undergoing adiabatic losses before emerging at the photosphere. The Arnett model\cite{Arnett1982} is often employed to fit SN light curves and to estimate key physical parameters of the ejecta and energy input. A commonly used expression for the bolometric luminosity at the photosphere, as introduced by Ref. \cite{Arnett1982}, is given by
\begin{equation}
L_{\mathrm{bol}}(t)=\frac{2L_{\mathrm{heat}0}}{\tau_d}e^{-\left(\frac{t^2+2t_\mathrm{ex}t}{\tau_d^2}\right)}\times\int_0^t\left(\frac{t_\mathrm{ex}+t^{\prime}}{\tau_d}\right)e^{\left(\frac{t^{\prime2}+2t_\mathrm{ex}t^{\prime}}{\tau_d^2}\right)}f_\mathrm{heat}(t^{\prime})dt^{\prime}+L_{\mathrm{Th}0}e^{-\left(\frac{t^2+2t_\mathrm{ex}t}{\tau_d^2}\right)}\,,
\end{equation}
where $L_{\rm{heat} 0}$ is the normalization constant for the heating source, representing the initial power from mechanisms such as radioactive decay or magnetar spindown. Here, $f_{\rm{heat}}(t)$ is the dimensionless time-dependent heating function that characterizes the evolution of the energy input, and $L_{\rm{Th} 0}$ denotes the initial cooling luminosity associated with the thermal energy deposited by the shock at early phases. The second term in the equation accounts for this shock-deposited thermal component. The expansion timescale is defined as $t_{\rm{ex}} = R_0 / v_{\rm{sc}}$, where $R_0$ is the initial radius of the ejecta and $v_{\rm{sc}}$ is the characteristic expansion velocity. The effective diffusion timescale $\tau_d$ depends on the opacity, ejecta mass, and expansion velocity, and is given by
\begin{equation}
    \tau_d = \sqrt{\frac{2 \kappa M_{\rm{ej}}}{\beta c v_{\rm{sc}}}}\, ,
\end{equation}
where $\beta \approx 13.8$ \cite{Arnett1982}.

The radioactive-decay-powered model yields an ejecta mass of \(M_{\text{ej}} = 2.38 \pm 0.06\, M_{\odot}\) and a nickel mass of \(M_{\text{Ni}} = 0.77 \pm 0.10\, M_{\odot}\) for SN\,2025kg, implying a high \(M_{\text{Ni}} / M_{\text{ej}}\) ratio. The fit is shown in the Extended Data Figure~\ref{fig:shock_cooling}. The kinetic energy is estimated to be \(E_{\text{K}} = 1.18 ^{+0.25}_{-0.32} \times 10^{52}\,\text{erg}\), suggesting rapid expansion driven by an energetic explosion. The \(\,^{56}\text{Ni}\) mass exceeds that of typical SNe~Ib/c, including GRB-associated events.

An alternative scenario involving rotational energy loss via magnetic dipole radiation was explored\cite{Kumar2024}, linking the magnetar's spindown rate to the radiative diffusion timescale. Fitting the light curve of SN\,2025kg with a magnetar-powered model yields best-fitting parameters of \(M_{\text{ej}} = 2.42^{+0.67}_{-0.70}\, M_{\odot}\), initial spin period \(P = 14.46 \pm 0.12\,\mathrm{ms}\), magnetic field strength \(B = (2.56 \pm 0.06) \times 10^{14}\,\mathrm{G}\), and kinetic energy \(E_{\text{K}} = 1.4^{+0.39}_{-0.42} \times 10^{52}\,\mathrm{erg}\); the fit is shown in Fig.~\ref{fig:slowjet}. With these parameters, the magnetar-powered model reproduces the observed light curve equally well as the radioactive-decay-powered model.

\subsection{Event rate estimation.\\}
We estimated the event rate density of fast X-ray transients similar to EP250108a following the methodology outlined in Refs. \cite{sun2015} and \cite{sun2022}, based on the first-year operational data from the EP. The EP-WXT has a field of view ($\Omega_{\rm WXT}$) of 3600 square degrees. By the time of the detection of EP250108a, WXT had accumulated approximately 1\,yr of operational time ($T_{\rm OT}$) with an observational duty cycle of $\eta=50\%$. Our simulation with WXT shows that EP250108a could be detected up to a maximum redshift of $z_{\rm max} = 0.19$ at a signal-to-noise ratio of 7. Within this redshift range, the effective maximum volume $V_{\rm max}$, weighted by the redshift distribution adopting the star formation history, is approximately $\sim 3.02$ $\rm Gpc^3$. 

The local event rate density of EP250108a is thus derived as
\begin{equation}
\rho_{0} \approx  \frac{4\pi}{\eta\Omega_{\rm WXT} T_{\rm OT} } \frac{1}{V_{\rm max}} \approx 7.3^{+16.8}_{-6.0} \,\rm Gpc^{-3} \, yr^{-1}\, ,
\end{equation}
where the 1$\sigma$ uncertainties are derived from small-sample statistics\cite{Gehrels1986ApJ}. This is estimated for one detection over the 1\,yr operation of EP-WXT. Accounting for the fact that only about 25\% of high-confidence FXTs have measured redshifts, we obtain a corrected event rate density of $\sim 29.2^{+67.2}_{-24.0}$ $\rm Gpc^{-3}\, yr^{-1}$. This value is comparable to the rate of LL-GRBs with luminosities similar to that of XRF\,060218\cite{sun2015}.

\clearpage

\section*{Data Availability}
The data presented in this work can be used or obtained at reasonable request. Some data are also publicly available through the corresponding references, GCN Circulars, or the UK Swift Science Data Centre website.

\section*{Code Availability}
The code (mainly in Python) used to produce the results and the figures will be available based on reasonable requests. Some public tools can be obtained as follows: the Source Extractor code is available at \url{https://astromatic.github.io/sextractor/}, the emcee Python package is available at \url{https://emcee.readthedocs.io/en/stable/}, the IRAF code at \url{https://iraf-community.github.io/}, the astrometry.net code at \url{https://astrometry.net/}, the PySALT package at \url{https://pysalt.salt.ac.za/}, the SciPy package at \url{https://scipy.org/}, the PSFex at \url{https://www.astromatic.net/software/psfex/}, the BANZAI pipline at \url{https://github.com/LCOGT/banzai}, the Xspec package at \url{https://heasarc.gsfc.nasa.gov/xanadu/xspec/}, and the SWarp code at \url{https://www.astromatic.net/software/swarp/}.

\bigskip
\bigskip
\bigskip
\bibliography{main}

\begin{addendum}

\item[Acknowledgments] 
This work is based on data obtained with the Einstein Probe (EP), a space mission supported by the Strategic Priority Program on Space Science of the Chinese Academy of Sciences, in collaboration with ESA, MPE, and CNES (grant XDA15310000) the Strategic Priority Program on Space Science of the Chinese Academy of Sciences (grant E02212A02S), and the Strategic Priority Research Program of the Chinese Academy of Sciences (grant XDB0550200). We acknowledge support by the National Natural Science Foundation of China (NSFC grants 12288102, 12373040, 12021003, 12103065, 12333004, 12203071, 12033003, 12233002, and 12303047).

This work makes use of data obtained with Mephisto, which is developed at and operated by the South-Western Institute for Astronomy Research of Yunnan University (SWIFAR-YNU), funded by the ``Yunnan University Development Plan for World-Class University'' and the ``Yunnan University Development Plan for World-Class Astronomy Discipline.'' The authors acknowledge additional support from the ``Science \& Technology Champion Project'' (202005AB160002) and from two ``Team Projects'' --- the ``Innovation Team'' (202105AE160021) and the ``Top Team'' (202305AT350002), all funded by the ``Yunnan Revitalization Talent Support Program.'' This work is further supported by the National Key Research and Development Program of China (2024YFA1611603).

A.V.F.’s group at UC Berkeley received financial assistance from the Christopher R. Redlich Fund, as well as donations from Gary and Cynthia Bengier, Clark and Sharon Winslow, Alan Eustace, William Draper, Timothy and Melissa Draper, Briggs and Kathleen Wood, Sanford Robertson (W.Z. is a Bengier-Winslow-Eustace Specialist in Astronomy, T.G.B. is a Draper-Wood-Robertson Specialist in Astronomy, Y.Y. was a Bengier-Winslow-Robertson Fellow in Astronomy), and numerous other donors.
A.V.F. is grateful for the
hospitality of the Hagler Institute for Advanced Study as well
as the Department of Physics and Astronomy at Texas A\&M
University during part of this investigation.

Some of the data presented herein were obtained at the W. M. Keck
Observatory, which is operated as a scientific partnership among the
California Institute of Technology, the University of California, and
NASA; the observatory was made possible by the generous financial
support of the W. M. Keck Foundation.
We acknowledge the support of the staff of the Keck~I 10\,m telescope and the 10.4\,m Gran Telescopio Canarias (GTC). 

This work makes use of the Las Cumbres Observatory global network of robotic telescopes. The LCO group is supported by NSF grants AST-1911225 and AST-1911151. S.B. and N. Elias-Rosa acknowledge support from the PRIN-INAF 2022, ``Shedding light on the nature of gap transients: from the observations to the models.'' 
We also use data supplied by the UK Swift Science Data Centre at the University of Leicester.

We gratefully acknowledge the China National Astronomical Data Center (NADC), the Astronomical Data Center of the Chinese Academy of Sciences, and the Chinese Virtual Observatory (China-VO) for providing data resources and technical support.

The work is supported by the National Key R\&D Program of China (grants 2022YFA1602902, 2023YFA1608100, 2023YFA1607800, and 2023YFA1607804), the National Natural Science Foundation of China (NSFC; grants 12120101003, 12373010, 12173051, and 12233008), and the Strategic Priority Research Program of the Chinese Academy of Sciences with grants XDB0550100 and XDB0550000.

\item[Author Contributions.] 
W.M.Y. has been leading the Einstein Probe project as Principal Investigator since the mission proposal stage. 
D.X., X.F. Wu, B.Z., and W.M.Y. initiated the study. 
D.X., H.S., X.F. Wang, Y.W.Y., W.X.L., Z.P.Z., X.Z.Z., J.J.G., X.F. Wu, and B.Z. ang coordinated the scientific investigations of the event and led the discussions.
S.Q.J., H.S., D.Y.L., and J.W.H. processed and analysed the WXT and FXT data, and simulated the behavior of XRF\,060218 in the WXT detection. S.Q.J. contributed to the {\it Fermi}/GBM data analysis. R.Z.L. is the transient advocate and contributed to the discovery and preliminary analysis of the event.
W.X.L., Z.P.Z., X.Z.Z., D.X., and X.F. Wang led the optical data taking and analysis. J.A., X.L., Z.P.Z., and X.Z.Z. contributed to optical data photometry.
B.M., E.H., D.F.X., M.G., C.-C.J., and J.L.D. contributed to obtaining and processing the Magellan spectra.
D.A.H.B., I.M., J.C., and L.F.W. contributed to obtaining and processing the SALT spectra.
C.R. Bom, C.D.K., L.S., A.S., M.G., and A.J.C. contributed to obtaining and processing the SOAR spectra and photometry.
A.V.F., Y.Y., T.G.B., W.Z., M.M.K., and W.J. contributed to obtaining and processing the Keck spectra.
Z.P.Z., X.L., D.X., and J.P.U.F. contributed to obtaining and processing the NOT optical spectroscopy and photometry.
D.A., N.C.S., F.P., A.L., and I.P. contributed to obtaining and processing the GTC spectra. S.Y.Y. helped reduce the GTC data.
X.Z.Z., Y.P.Y., R.Z.L., J.R.M., L.P.L., X.W.L., J.H.Z., G.W.D., Y.F., B.K., K.C., contributed to the Mephisto data taking and photometry.
W.X.L., M.A., J.F., D.A.H., M.N., E.P.G., C.M., and G.T. contributed to the LCO network data taking.
A.M., O.S., O.M., V.G., S.B.P., N.E., M.J., K.A., L.Z.W., L.X.Y., Z.G., G.P., J.J.Z., Z.Y.W., L.P.L., Y.L.H, B.T.W., and Y.D.H. contributed to the optical data taking and photometry.
T.A. and Y.Q.L. helped with the radio data taking and analysis. 
B.Z., D.X., L.X., S.Q.J., and Z.P.Z. contributed to comparing this event with GRBs.
L.W.X., Z.P.Z., J.A., D.X., and B.Z. contributed to comparing this event with SNe.
H.S. contributed to the event rate density. 
B.Z., J.J.G., L.D.L., Y.H.W., and Y.W.Y. led the theoretical investigation of the event. X.F. Wu, Y.H.Z., G.L.W., D.X., H.S., W.X.L., and Z.P.Z. contributed to the theoretical investigation of the event.
Y.L., C.Z., Z.X.L., H.Q.C., H.W.P., Y.F.X., Y.C., S.M.J., C.K.L., X.J.S., and Y.H.Z. contributed to the development of the WXT and FXT instruments, calibration data, or data-analysis software.
W.X.L., Z.P.Z., D.X., X.Z.Z., J.J.G., L.D.L., H.S., S.Q.J., J.W.H., and T.A. drafted the manuscript with the help of all authors. A.V.F. edited the manuscript. 

\item[Competing Interests] The authors declare that they have no competing financial interests.

\end{addendum}

\setcounter{figure}{0}
\setcounter{table}{0}
\captionsetup[figure]{labelfont={bf}, labelformat={default}, labelsep=period, name={Extended Data Fig.}}
\captionsetup[table]{labelfont={bf}, labelformat={default}, labelsep=period, name={Extended Data Table}}


\clearpage

\begin{figure}
\centering
\begin{overpic}[width=0.4\textwidth]{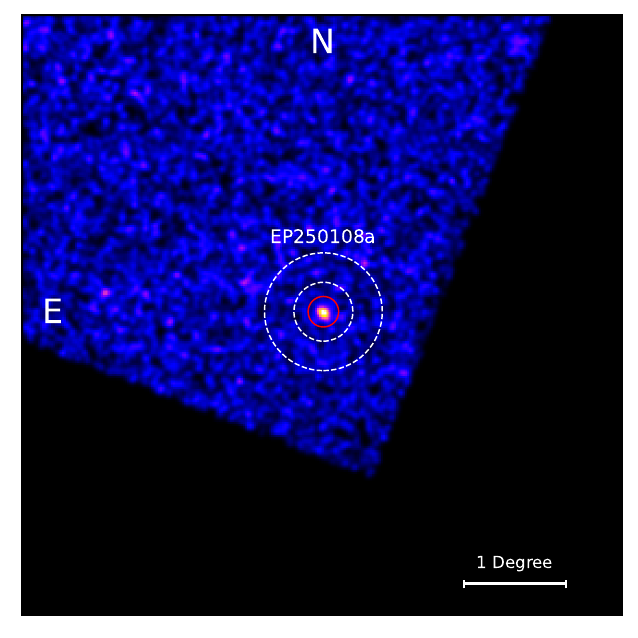}\put(-5, 106){\bf a}\end{overpic} 
\begin{overpic}[width=0.55\textwidth]{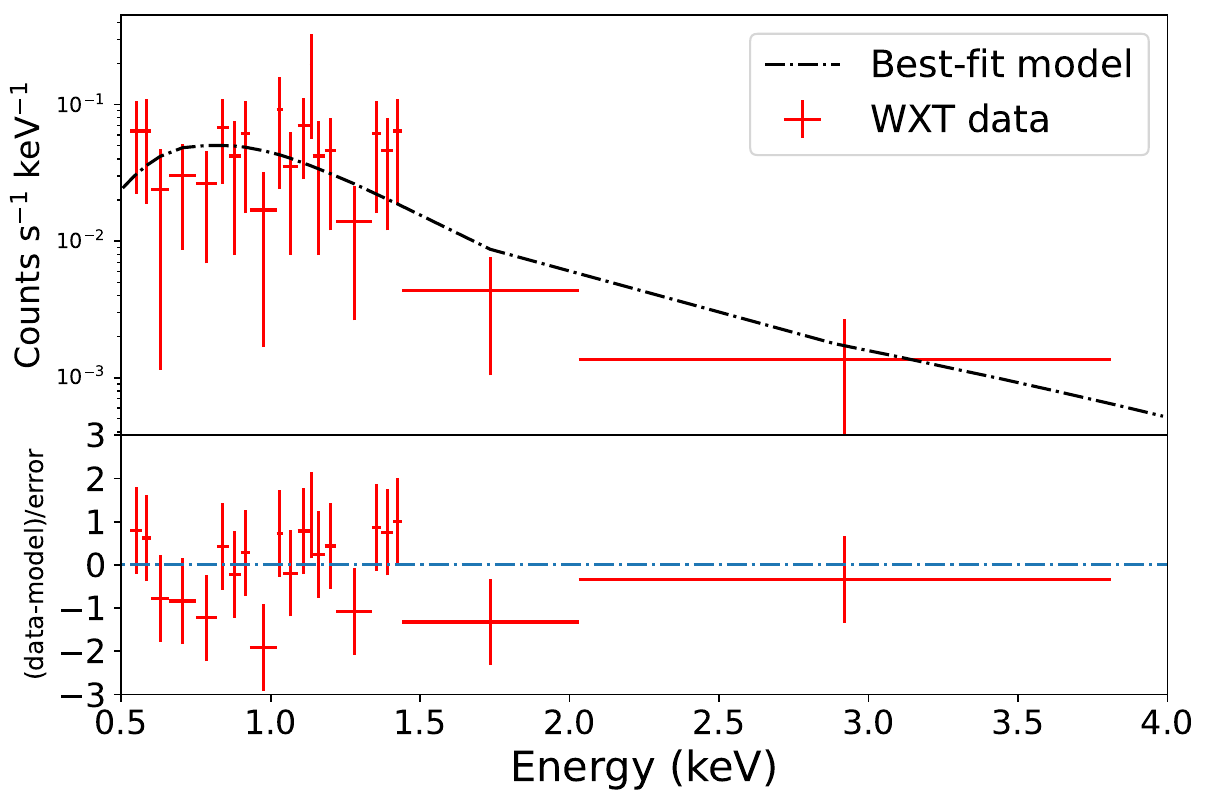}\put(1, 76){\bf b}\end{overpic}
\caption{\noindent\textbf{WXT image and spectrum.} 
\textbf{a}, The location plot of EP250108a. The red circle represents the $9'$ aperture used for extracting the light curve and spectrum of EP250108a. The white annulus with radii of $18'$ and $36'$ indicates the background region.
\textbf{b}, The WXT spectrum extracted between the $T_{90}$ interval and from the source region illustrated in the left panel. The upper panel shows the observed WXT spectrum and its best-fit model. The bottom panel shows the statistical value for each data point. The blue dashed line represents the zero value.
}
\label{fig:X_ray_detail}
\end{figure}
\clearpage

\begin{figure}
\centering
\begin{tabular}{c}
\includegraphics[width=0.85\textwidth]{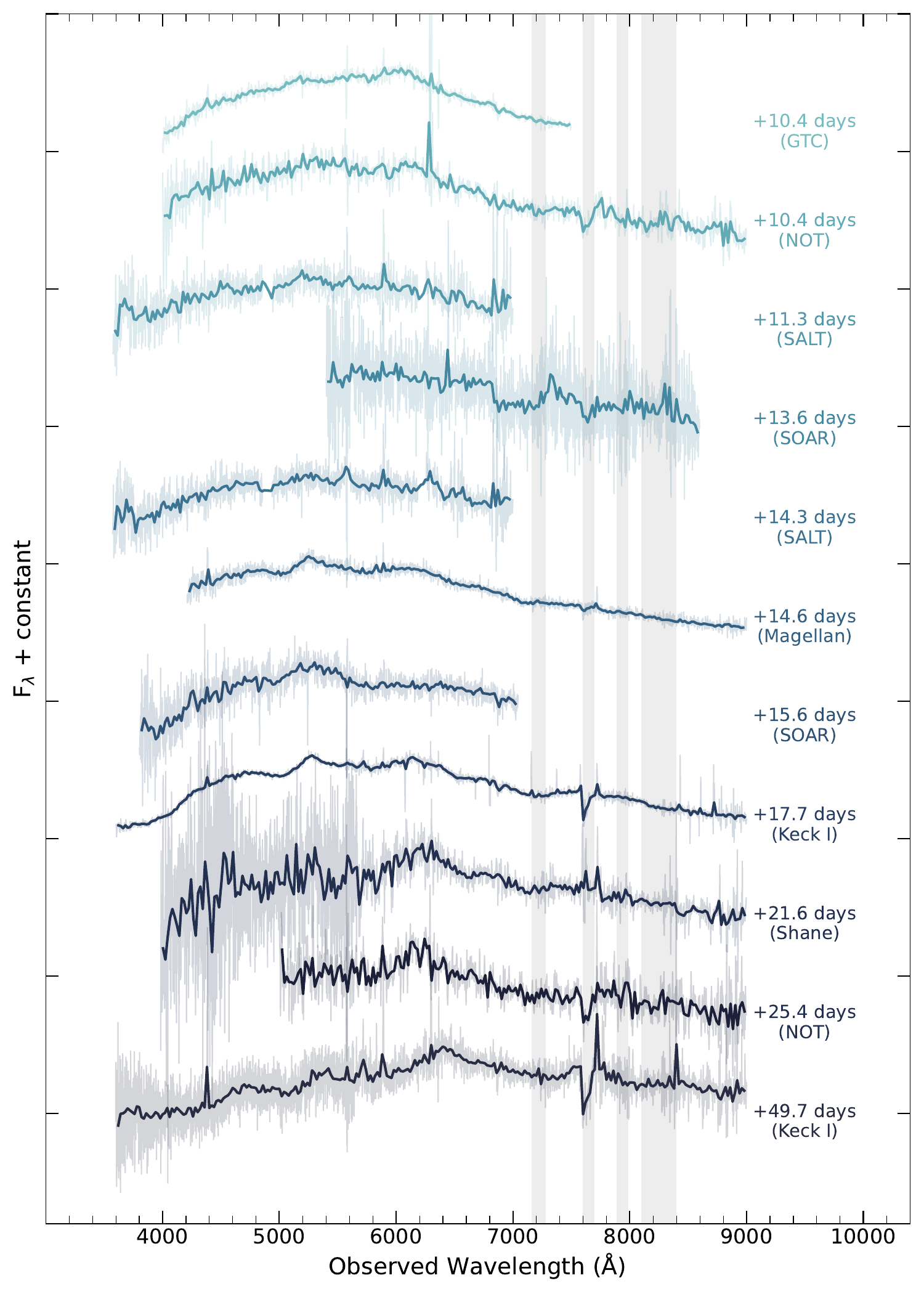}
\end{tabular}
\caption{\noindent\textbf{Optical spectra evolution of SN\,2025kg.} The observed spectra are arranged in chronological order from top to bottom. The observation time is labeled on the right side with the same color as the corresponding spectrum. All of the spectra are rebinned with a width of 20\,\AA ~ for display purposes, while the original spectra are plotted in lighter colors. The grey regions show the bands affected by telluric absorption.
}
\label{fig:opt_sp}
\end{figure} 
\clearpage

\begin{figure}
\centering
\includegraphics[width=0.8\textwidth]{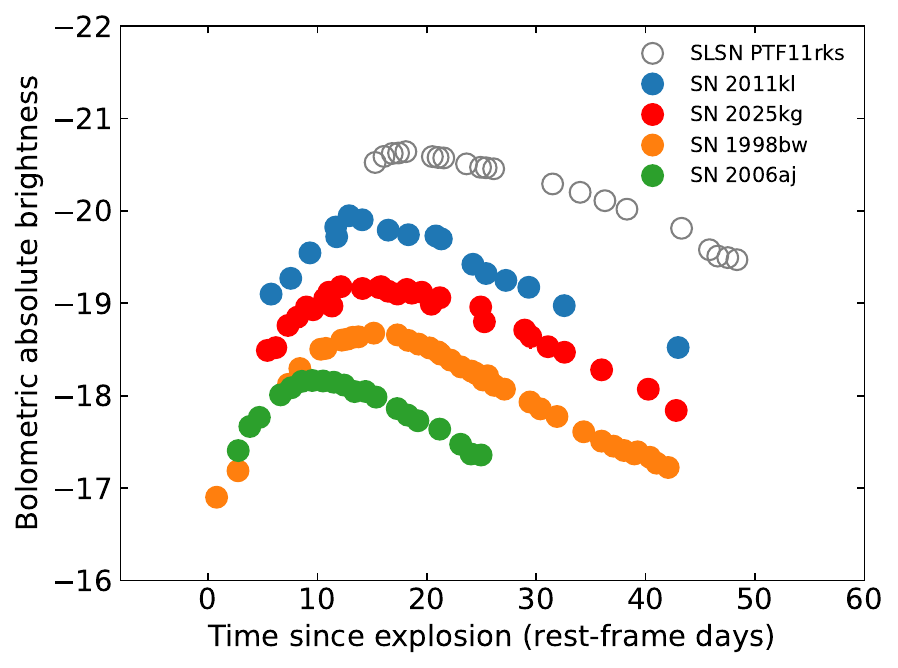}
\caption{\noindent\textbf{The pseudobolometric light curve of SN\,2025kg.} The light curve corresponds to rest-frame wavelengths of 3000--9000\,\AA, and is compared with those of representative GRB- or XRF-associated supernovae, including GRB\,980425/SN\,1998bw\cite{1998Natur.395..670G}, XRF\,060218/SN\,2006aj\cite{2006Natur.442.1018M}, and GRB\,111209A/SN\,2011kl\cite{2015Natur.523..189G}. The light curve of the hydrogen-poor superluminous supernova PTF11rks\cite{2013ApJ...770..128I} is also shown, highlighting the distinction between these events.}
\label{fig:bolometric}
\end{figure}

\begin{figure}
\centering
\includegraphics[width=0.8\textwidth]{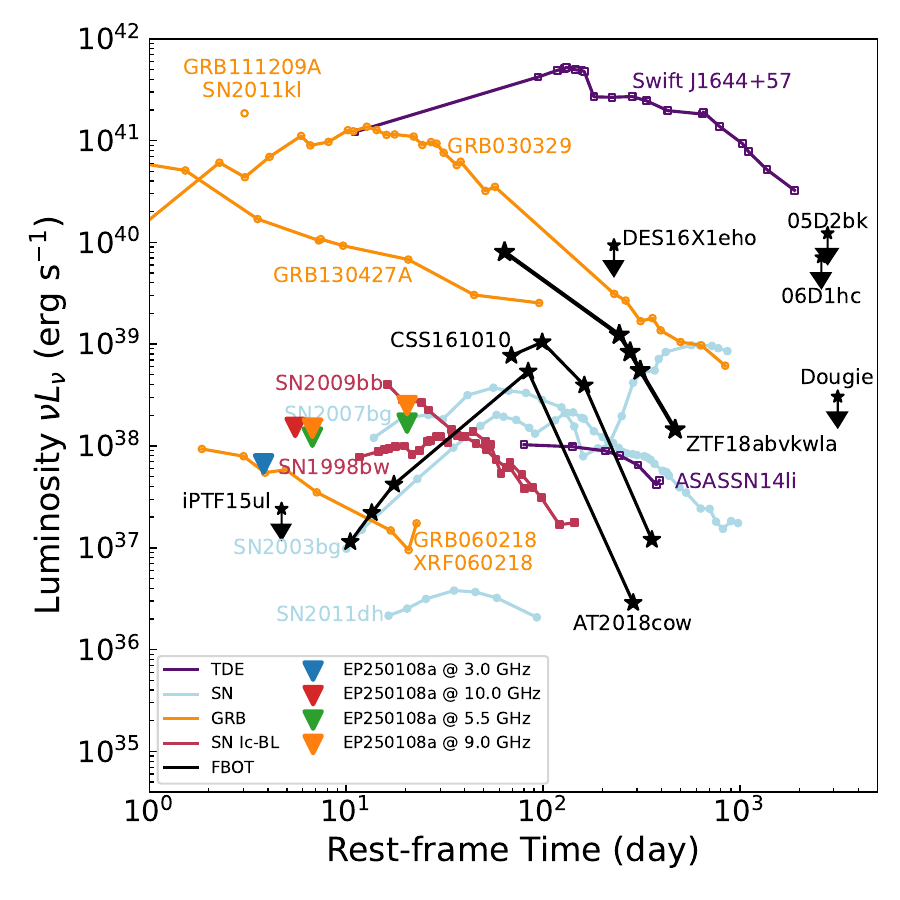}
\caption{\noindent\textbf{Radio luminosity upper limits of EP250108a/SN\,2025kg.} The 3\,GHz, 10\,GHz, 5.5\,GHz, and 9\,GHz upper limits of EP250108a/SN\,2025kg are compared to low-frequency (1–-10\,GHz) light curves of different classes of energetic explosions: tidal disruption events\cite{Berger12,Alexander16}, SNe\cite{Soderberg05,Salas13}, relativistic SNe~Ic-BL\cite{Kulkarni98,Soderberg10}, long-duration GRBs\cite{Berger03,Soderberg06,Perley14}, and fast blue optical transients\cite{Margutti19,Coppejans20,Ho20}. }
\label{fig:radio_LC}
\end{figure}

\clearpage

\begin{figure}
\centering
\begin{overpic}[width=0.9\textwidth]{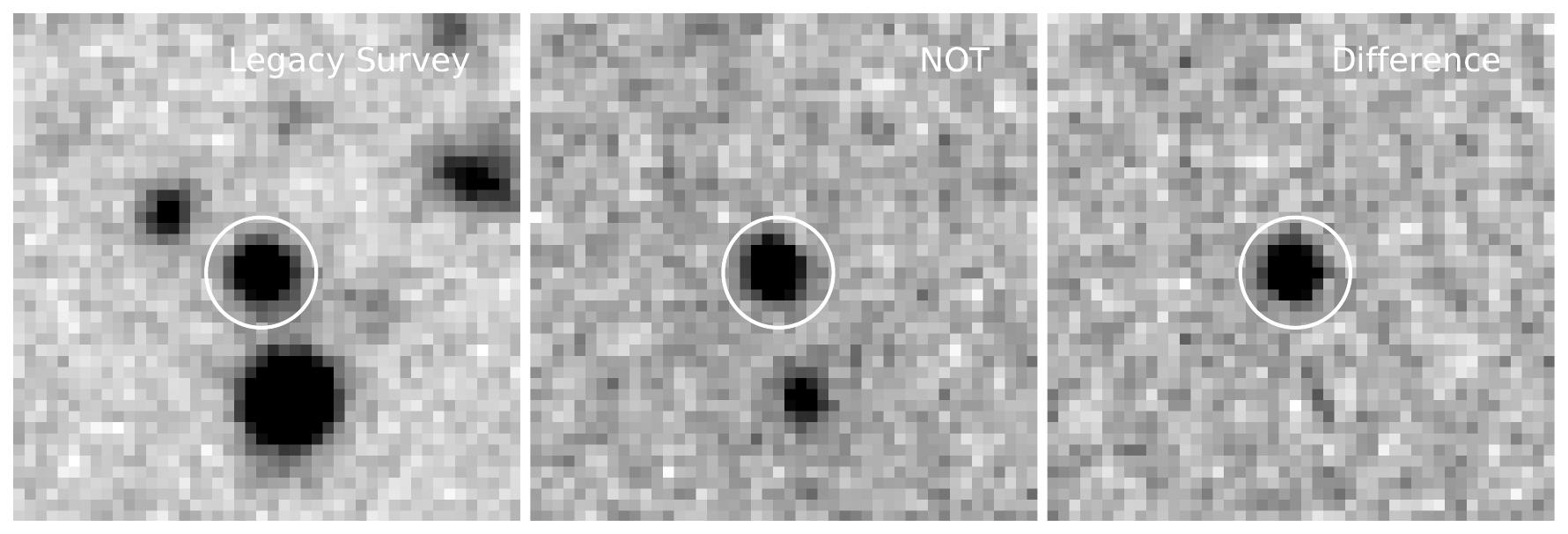}\put(-2, 36){\bf a}\end{overpic} 
\begin{overpic}[width=0.9\textwidth]{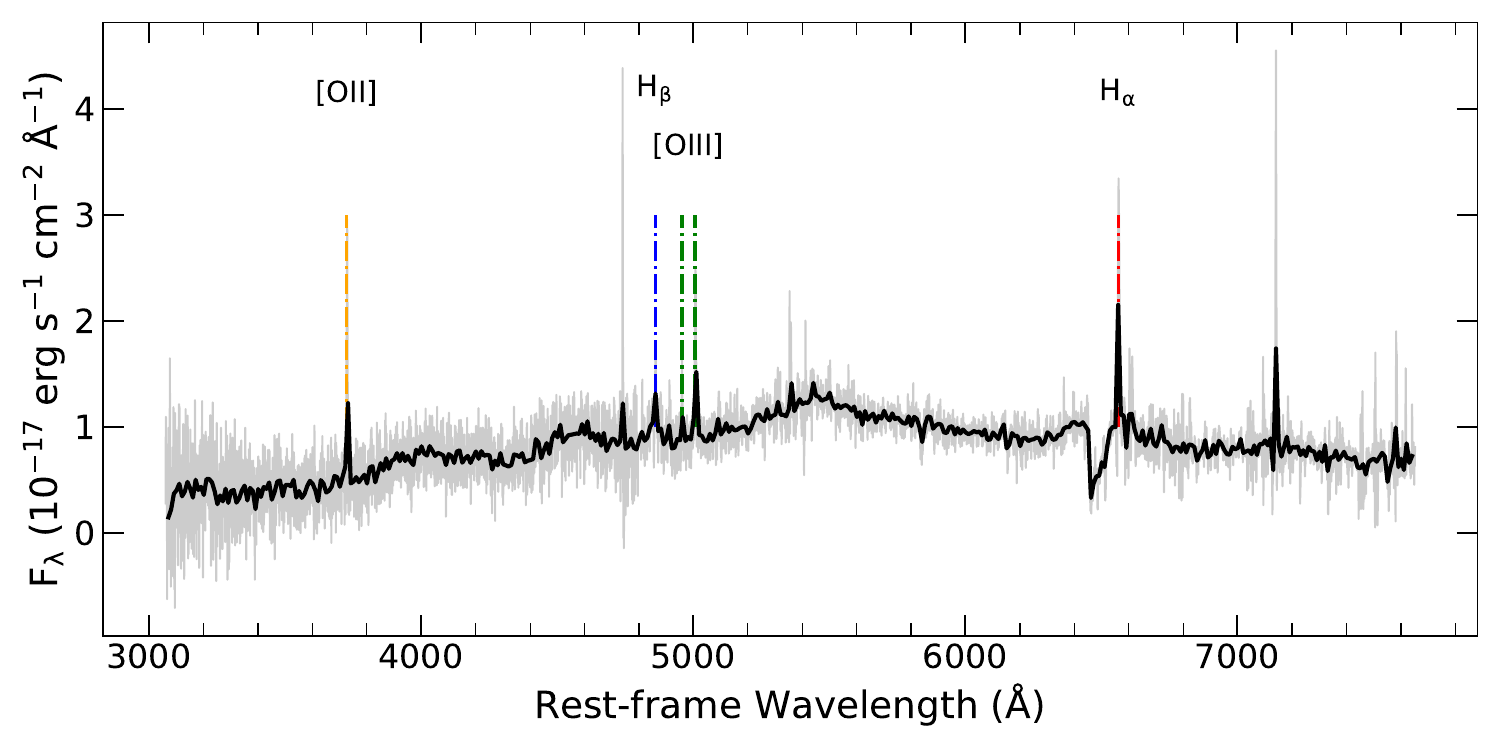}\put(-2, 56){\bf b}\end{overpic}
\caption{\noindent\textbf{The host galaxy of EP250108a/SN\,2025kg.} 
\textbf{a}, The location plot of EP250108a/SN\,2025kg. From left to right, the shown images are taken from the Legacy Survey archive, the NOT first-night observation, and the subtracted residual image. 
\textbf{b}, The optical spectrum obtained with Keck/LRIS on 26 February shown with a rebinned width of 10~\AA, together with the light-colored raw spectrum. Prominent narrow emission lines from the host galaxy are marked with vertical dashed lines and color-coded by line species, including [O~II] $\lambda3727$, H$\beta$, and [O~III] $\lambda\lambda4959$, 5007, as well as H$\alpha$. The spectral continuum is still dominated by SN\,2025kg.
}
\label{fig:host}
\end{figure}
\clearpage

\begin{figure}
\centering
\begin{tabular}{c}
\includegraphics[width=0.8\textwidth]{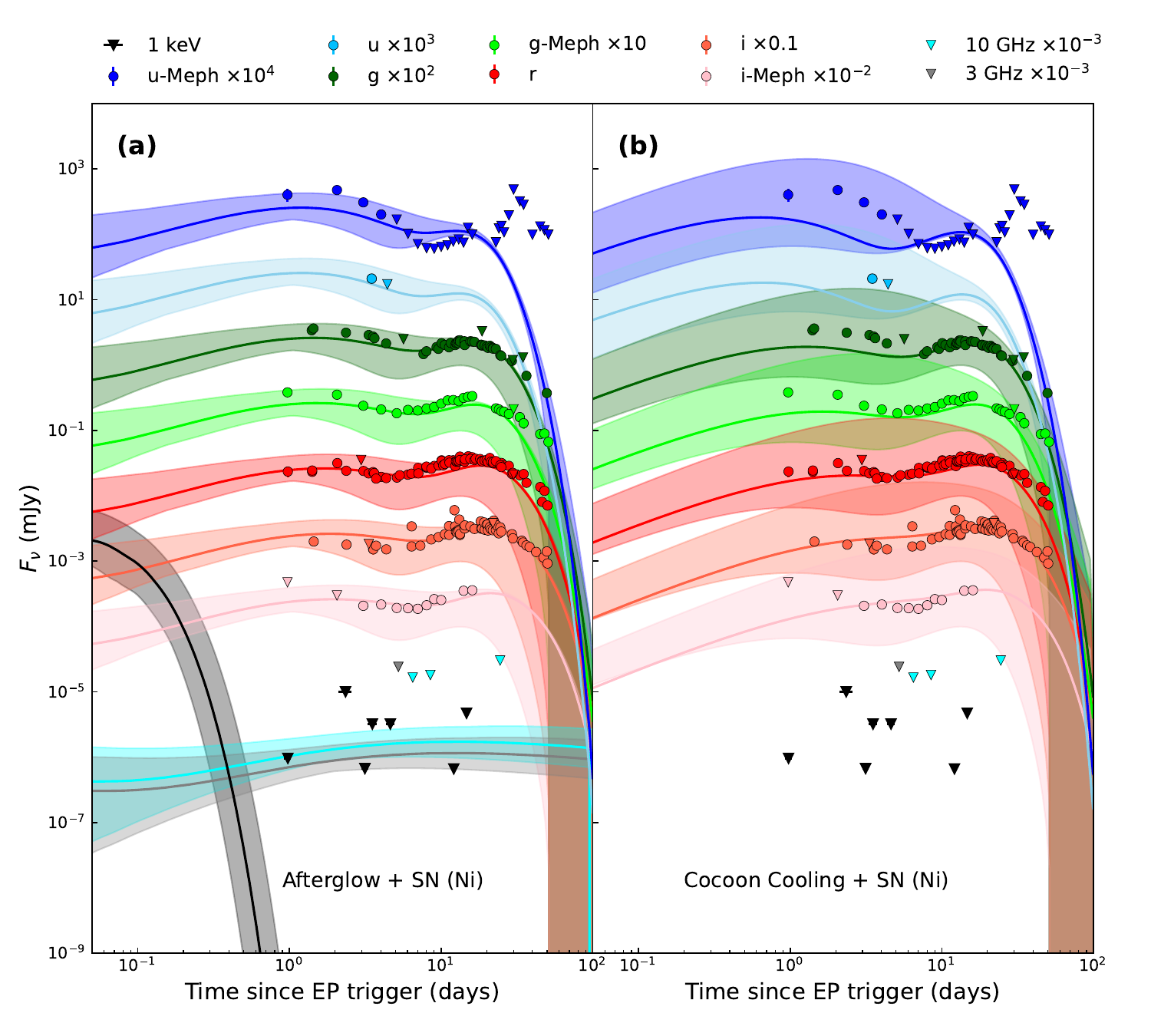}
\end{tabular}
\caption{\noindent\textbf{Radioactive-decay-powered modeling of EP250108a.} The early emission is attributed to (a) the jet and (b) the cooling of the cocoon material (b).
The late-time supernova emission is described by the radioactive Ni.}
\label{fig:shock_cooling}
\end{figure}

\clearpage

\begin{longtable}{ccccc}
\caption{Optical photometry of SN\,2025kg. Errors represent the 1$\sigma$ uncertainties. The \textit{ugriz}-band photometry is given in the AB magnitude system, while all other bands are in the Vega magnitude system.} \label{tab:opt_ph} \\ 
\toprule
$\Delta\,T$ (day) & Mag  & Filter & Telescope & Ref.\\ 
\midrule
\endfirsthead
\toprule
$\Delta\,T$ (day) & Mag  & Filter & Telescope  & Ref. \\ 
\midrule
\endhead
\midrule
\multicolumn{5}{c}{{Continued on next page}} \\ 
\midrule
\endfoot
\bottomrule
\endlastfoot
0.97 & 20.0 $\pm$ 0.25 & $ u $ & Mephisto & This work \\
2.06 & 19.82 $\pm$ 0.15 & $ u $ & Mephisto & This work \\
3.07 & 20.29 $\pm$ 0.14 & $ u $ & Mephisto & This work \\
4.03 & 20.76 $\pm$ 0.15 & $ u $ & Mephisto & This work \\
0.98 & 20.03 $\pm$ 0.19 & $ v $ & Mephisto & This work \\
2.07 & 20.56 $\pm$ 0.21 & $ v $ & Mephisto & This work \\
3.08 & 20.41 $\pm$ 0.13 & $ v $ & Mephisto & This work \\
4.03 & 20.87 $\pm$ 0.16 & $ v $ & Mephisto & This work \\
0.97 & 20.02 $\pm$ 0.15 & $ g $ & Mephisto & This work \\
2.06 & 20.11 $\pm$ 0.16 & $ g $ & Mephisto & This work \\
3.07 & 20.53 $\pm$ 0.09 & $ g $ & Mephisto & This work \\
4.03 & 20.68 $\pm$ 0.08 & $ g $ & Mephisto & This work \\
5.10 & 20.82 $\pm$ 0.09 & $ g $ & Mephisto & This work \\
6.05 & 20.69 $\pm$ 0.06 & $ g $ & Mephisto & This work \\
7.02 & 20.72 $\pm$ 0.04 & $ g $ & Mephisto & This work \\
8.03 & 20.64 $\pm$ 0.04 & $ g $ & Mephisto & This work \\
9.00 & 20.58 $\pm$ 0.04 & $ g $ & Mephisto & This work \\
10.02 & 20.45 $\pm$ 0.03 & $ g $ & Mephisto & This work \\
11.05 & 20.33 $\pm$ 0.04 & $ g $ & Mephisto & This work \\
12.04 & 20.31 $\pm$ 0.03 & $ g $ & Mephisto & This work \\
13.07 & 20.36 $\pm$ 0.04 & $ g $ & Mephisto & This work \\
14.07 & 20.23 $\pm$ 0.03 & $ g $ & Mephisto & This work \\
15.04 & 20.19 $\pm$ 0.04 & $ g $ & Mephisto & This work \\
16.05 & 20.16 $\pm$ 0.03 & $ g $ & Mephisto & This work \\
23.00 & 20.63 $\pm$ 0.05 & $ g $ & Mephisto & This work \\
24.06 & 20.67 $\pm$ 0.08 & $ g $ & Mephisto & This work \\
25.00 & 20.75 $\pm$ 0.09 & $ g $ & Mephisto & This work \\
26.01 & 20.76 $\pm$ 0.07 & $ g $ & Mephisto & This work \\
28.01 & 20.85 $\pm$ 0.13 & $ g $ & Mephisto & This work \\
33.05 & 20.97 $\pm$ 0.18 & $ g $ & Mephisto & This work \\
35.00 & 21.2 $\pm$ 0.2 & $ g $ & Mephisto & This work \\
44.99 & 21.62 $\pm$ 0.17 & $ g $ & Mephisto & This work \\
48.03 & 21.6 $\pm$ 0.14 & $ g $ & Mephisto & This work \\
51.00 & 21.92 $\pm$ 0.2 & $ g $ & Mephisto & This work \\
0.98 & 20.54 $\pm$ 0.2 & $ r $ & Mephisto & This work \\
2.07 & 20.22 $\pm$ 0.09 & $ r $ & Mephisto & This work \\
3.08 & 20.52 $\pm$ 0.08 & $ r $ & Mephisto & This work \\
4.03 & 20.74 $\pm$ 0.07 & $ r $ & Mephisto & This work \\
5.12 & 20.77 $\pm$ 0.08 & $ r $ & Mephisto & This work \\
6.03 & 20.66 $\pm$ 0.05 & $ r $ & Mephisto & This work \\
7.06 & 20.4 $\pm$ 0.04 & $ r $ & Mephisto & This work \\
8.00 & 20.39 $\pm$ 0.03 & $ r $ & Mephisto & This work \\
9.03 & 20.32 $\pm$ 0.03 & $ r $ & Mephisto & This work \\
10.00 & 20.25 $\pm$ 0.03 & $ r $ & Mephisto & This work \\
11.07 & 20.18 $\pm$ 0.03 & $ r $ & Mephisto & This work \\
12.02 & 20.11 $\pm$ 0.03 & $ r $ & Mephisto & This work \\
13.09 & 20.14 $\pm$ 0.04 & $ r $ & Mephisto & This work \\
14.10 & 20.07 $\pm$ 0.03 & $ r $ & Mephisto & This work \\
15.11 & 19.96 $\pm$ 0.05 & $ r $ & Mephisto & This work \\
16.02 & 19.99 $\pm$ 0.02 & $ r $ & Mephisto & This work \\
23.03 & 20.2 $\pm$ 0.04 & $ r $ & Mephisto & This work \\
24.03 & 20.24 $\pm$ 0.04 & $ r $ & Mephisto & This work \\
25.02 & 20.21 $\pm$ 0.05 & $ r $ & Mephisto & This work \\
25.99 & 20.3 $\pm$ 0.04 & $ r $ & Mephisto & This work \\
27.99 & 20.32 $\pm$ 0.09 & $ r $ & Mephisto & This work \\
33.01 & 20.68 $\pm$ 0.08 & $ r $ & Mephisto & This work \\
35.03 & 20.63 $\pm$ 0.11 & $ r $ & Mephisto & This work \\
45.02 & 21.13 $\pm$ 0.08 & $ r $ & Mephisto & This work \\
48.01 & 21.28 $\pm$ 0.1 & $ r $ & Mephisto & This work \\
3.07 & 21.01 $\pm$ 0.14 & $ i $ & Mephisto & This work \\
4.03 & 20.96 $\pm$ 0.1 & $ i $ & Mephisto & This work \\
5.10 & 21.09 $\pm$ 0.12 & $ i $ & Mephisto & This work \\
6.05 & 21.1 $\pm$ 0.1 & $ i $ & Mephisto & This work \\
7.02 & 21.12 $\pm$ 0.08 & $ i $ & Mephisto & This work \\
8.03 & 20.98 $\pm$ 0.1 & $ i $ & Mephisto & This work \\
9.00 & 20.75 $\pm$ 0.09 & $ i $ & Mephisto & This work \\
10.02 & 20.78 $\pm$ 0.07 & $ i $ & Mephisto & This work \\
14.07 & 20.43 $\pm$ 0.06 & $ i $ & Mephisto & This work \\
16.05 & 20.41 $\pm$ 0.07 & $ i $ & Mephisto & This work \\

1.44 & 20.09 $\pm$ 0.02 & $g'$ & NOT & This work \\
2.36 & 20.25 $\pm$ 0.04 & $g'$ & NOT & This work \\
3.33 & 20.34 $\pm$ 0.18 & $g'$ & LT & This work \\
3.56 & 20.4 $\pm$ 0.13 & $g'$ & LCO & This work \\
3.64 & 20.46 $\pm$ 0.12 & $g'$ & LCO & This work \\
4.32 & 20.61 $\pm$ 0.07 & $g'$ & GTC & This work \\
4.34 & 20.66 $\pm$ 0.09 & $g'$ & NOT & This work \\
7.65 & 21.07 $\pm$ 0.15 & $g'$ & LCO & This work \\
8.00 & 20.96 $\pm$ 0.1 & $g'$ & LJT & This work \\
9.34 & 20.8 $\pm$ 0.05 & $g'$ & LCO & This work \\
9.60 & 20.87 $\pm$ 0.06 & $g'$ & LCO & This work \\
10.07 & 20.65 $\pm$ 0.05 & $g'$ & LJT & This work \\
10.58 & 20.72 $\pm$ 0.03 & $g'$ & LCO & This work \\
11.57 & 20.64 $\pm$ 0.03 & $g'$ & LCO & This work \\
12.53 & 20.65 $\pm$ 0.05 & $g'$ & LCO & This work \\
12.70 & 20.66 $\pm$ 0.03 & $g'$ & LCO & This work \\
13.00 & 20.65 $\pm$ 0.03 & $g'$ & LCO & This work \\
13.04 & 20.55 $\pm$ 0.04 & $g'$ & LCO & This work \\
13.29 & 20.54 $\pm$ 0.08 & $g'$ & SAAO-1\,m & This work \\
13.32 & 20.59 $\pm$ 0.04 & $g'$ & LCO & This work \\
14.28 & 20.58 $\pm$ 0.03 & $g'$ & LCO & This work \\
15.58 & 20.58 $\pm$ 0.04 & $g'$ & LCO & This work \\
15.99 & 20.59 $\pm$ 0.07 & $g'$ & LCO & This work \\
16.59 & 20.6 $\pm$ 0.03 & $g'$ & LCO & This work \\
18.38 & 20.73 $\pm$ 0.04 & $g'$ & LT & This work \\
18.40 & 20.71 $\pm$ 0.04 & $g'$ & LT & This work \\
18.59 & 20.74 $\pm$ 0.04 & $g'$ & LCO & This work \\
18.61 & 20.73 $\pm$ 0.05 & $g'$ & LCO & This work \\
19.38 & 20.74 $\pm$ 0.04 & $g'$ & LT & This work \\
19.96 & 20.79 $\pm$ 0.03 & $g'$ & LCO & This work \\
20.36 & 20.83 $\pm$ 0.04 & $g'$ & LT & This work \\
20.96 & 20.75 $\pm$ 0.05 & $g'$ & LCO & This work \\
21.37 & 20.81 $\pm$ 0.06 & $g'$ & LT & This work \\
21.93 & 20.83 $\pm$ 0.06 & $g'$ & LCO & This work \\
21.97 & 20.86 $\pm$ 0.05 & $g'$ & LCO & This work \\
22.96 & 20.87 $\pm$ 0.05 & $g'$ & LCO & This work \\
24.56 & 21.11 $\pm$ 0.18 & $g'$ & LCO & This work \\
24.93 & 21.14 $\pm$ 0.06 & $g'$ & LCO & This work \\
29.30 & 21.33 $\pm$ 0.15 & $g'$ & LCO & This work \\
36.53 & 21.9 $\pm$ 0.1 & $g'$ & SOAR & This work \\
49.77 & 22.56 $\pm$ 0.14 & $g'$ & Keck & This work \\
1.41 & 20.55 $\pm$ 0.03 & $r'$ & NOT & This work \\
1.41 & 20.5 $\pm$ 0.1 & $r'$ & LT & This work \\
2.34 & 20.46 $\pm$ 0.09 & $r'$ & CAHA-2.2\,m & This work \\
2.37 & 20.51 $\pm$ 0.06 & $r'$ & NOT & This work \\
3.33 & 20.61 $\pm$ 0.15 & $r'$ & LT & This work \\
3.56 & 20.56 $\pm$ 0.13 & $r'$ & LCO & This work \\
3.62 & 20.6 $\pm$ 0.1 & $r'$ & LCO & This work \\
4.33 & 20.7 $\pm$ 0.07 & $r'$ & GTC & This work \\
4.35 & 20.79 $\pm$ 0.09 & $r'$ & NOT & This work \\
5.34 & 20.68 $\pm$ 0.1 & $r'$ & LCO & This work \\
6.37 & 20.62 $\pm$ 0.05 & $r'$ & LT & This work \\
7.09 & 20.61 $\pm$ 0.12 & $r'$ & LJT & This work \\
8.01 & 20.35 $\pm$ 0.06 & $r'$ & LJT & This work \\
8.62 & 20.43 $\pm$ 0.07 & $r'$ & LCO & This work \\
9.63 & 20.31 $\pm$ 0.05 & $r'$ & LCO & This work \\
10.08 & 20.13 $\pm$ 0.04 & $r'$ & LJT & This work \\
10.61 & 20.25 $\pm$ 0.03 & $r'$ & LCO & This work \\
11.28 & 20.11 $\pm$ 0.07 & $r'$ & SAAO-1\,m & This work \\
11.36 & 20.2 $\pm$ 0.03 & $r'$ & LCO & This work \\
12.54 & 20.15 $\pm$ 0.03 & $r'$ & LCO & This work \\
12.70 & 20.16 $\pm$ 0.03 & $r'$ & LCO & This work \\
13.00 & 20.16 $\pm$ 0.03 & $r'$ & LCO & This work \\
13.04 & 20.14 $\pm$ 0.03 & $r'$ & LCO & This work \\
13.29 & 19.97 $\pm$ 0.07 & $r'$ & SAAO-1\,m & This work \\
13.33 & 20.1 $\pm$ 0.04 & $r'$ & LCO & This work \\
14.29 & 20.16 $\pm$ 0.03 & $r'$ & LCO & This work \\
14.31 & 20.07 $\pm$ 0.07 & $r'$ & SAAO-1\,m & This work \\
15.59 & 20.06 $\pm$ 0.03 & $r'$ & LCO & This work \\
16.60 & 20.1 $\pm$ 0.03 & $r'$ & LCO & This work \\
16.90 & 20.05 $\pm$ 0.04 & $r'$ & LCO & This work \\
18.60 & 20.13 $\pm$ 0.03 & $r'$ & LCO & This work \\
18.62 & 20.15 $\pm$ 0.05 & $r'$ & LCO & This work \\
19.34 & 20.12 $\pm$ 0.02 & $r'$ & LT & This work \\
19.96 & 20.10 $\pm$ 0.03 & $r'$ & LCO & This work \\
20.37 & 20.18 $\pm$ 0.02 & $r'$ & LT & This work \\
20.97 & 20.02 $\pm$ 0.04 & $r'$ & LCO & This work \\
21.37 & 20.16 $\pm$ 0.04 & $r'$ & LT & This work \\
21.93 & 20.22 $\pm$ 0.04 & $r'$ & LCO & This work \\
21.97 & 20.18 $\pm$ 0.04 & $r'$ & LCO & This work \\
22.97 & 20.14 $\pm$ 0.04 & $r'$ & LCO & This work \\
24.56 & 20.3 $\pm$ 0.07 & $r'$ & LCO & This work \\
24.93 & 20.37 $\pm$ 0.04 & $r'$ & LCO & This work \\
29.31 & 20.63 $\pm$ 0.09 & $r'$ & LCO & This work \\
29.71 & 20.59 $\pm$ 0.09 & $r'$ & LCO & This work \\
36.58 & 20.97 $\pm$ 0.04 & $r'$ & SOAR & This work \\
46.32 & 21.7 $\pm$ 0.11 & $r'$ & NOT & This work \\
50.32 & 21.84 $\pm$ 0.08 & $r'$ & NOT & This work \\
1.45 & 20.7 $\pm$ 0.04 & $i'$ & NOT & This work \\
2.38 & 20.82 $\pm$ 0.07 & $i'$ & NOT & This work \\
3.73 & 20.86 $\pm$ 0.11 & $i'$ & LCO & This work \\
4.33 & 20.92 $\pm$ 0.04 & $i'$ & GTC & This work \\
4.35 & 21.0 $\pm$ 0.07 & $i'$ & NOT & This work \\
6.39 & 20.89 $\pm$ 0.07 & $i'$ & LT & This work \\
7.30 & 20.86 $\pm$ 0.11 & $i'$ & LCO & This work \\
8.61 & 20.62 $\pm$ 0.09 & $i'$ & LCO & This work \\
9.62 & 20.53 $\pm$ 0.09 & $i'$ & LCO & This work \\
10.59 & 20.42 $\pm$ 0.05 & $i'$ & LCO & This work \\
11.30 & 20.45 $\pm$ 0.06 & $i'$ & LCO & This work \\
12.55 & 20.33 $\pm$ 0.06 & $i'$ & LCO & This work \\
12.70 & 20.34 $\pm$ 0.08 & $i'$ & LCO & This work \\
13.01 & 20.26 $\pm$ 0.05 & $i'$ & LCO & This work \\
13.04 & 20.31 $\pm$ 0.07 & $i'$ & LCO & This work \\
14.29 & 20.2 $\pm$ 0.05 & $i'$ & LCO & This work \\
15.60 & 20.12 $\pm$ 0.05 & $i'$ & LCO & This work \\
16.61 & 20.22 $\pm$ 0.05 & $i'$ & LCO & This work \\
18.41 & 20.21 $\pm$ 0.03 & $i'$ & LT & This work \\
18.61 & 20.2 $\pm$ 0.05 & $i'$ & LCO & This work \\
19.39 & 20.25 $\pm$ 0.03 & $i'$ & LT & This work \\
20.37 & 20.28 $\pm$ 0.03 & $i'$ & LT & This work \\
21.38 & 20.23 $\pm$ 0.05 & $i'$ & LT & This work \\
21.93 & 20.27 $\pm$ 0.07 & $i'$ & LCO & This work \\
22.98 & 20.26 $\pm$ 0.08 & $i'$ & LCO & This work \\
24.01 & 20.39 $\pm$ 0.06 & $i'$ & LJT & This work \\
24.93 & 20.32 $\pm$ 0.06 & $i'$ & LCO & This work \\
29.32 & 20.42 $\pm$ 0.11 & $i'$ & LCO & This work \\
29.71 & 20.58 $\pm$ 0.12 & $i'$ & LCO & This work \\
34.05 & 20.67 $\pm$ 0.1 & $i'$ & LJT & This work \\
34.70 & 20.74 $\pm$ 0.13 & $i'$ & LCO & This work \\
36.55 & 20.85 $\pm$ 0.04 & $i'$ & SOAR & This work \\
38.32 & 20.91 $\pm$ 0.08 & $i'$ & NOT & This work \\
42.31 & 21.1 $\pm$ 0.05 & $i'$ & NOT & This work \\
47.33 & 21.31 $\pm$ 0.05 & $i'$ & NOT & This work \\
50.33 & 21.54 $\pm$ 0.08 & $i'$ & NOT & This work \\
1.43 & 20.96 $\pm$ 0.06 & $z'$ & NOT & This work \\
2.38 & 21.04 $\pm$ 0.18 & $z'$ & NOT & This work \\
3.73 & 21.12 $\pm$ 0.15 & $z'$ & LCO & This work \\
4.33 & 21.17 $\pm$ 0.09 & $z'$ & GTC & This work \\
12.70 & 20.51 $\pm$ 0.08 & $z'$ & LCO & This work \\
13.04 & 20.61 $\pm$ 0.10 & $z'$ & LCO & This work \\
19.34 & 20.53 $\pm$ 0.06 & $z'$ & LT & This work \\
20.38 & 20.36 $\pm$ 0.08 & $z'$ & LT & This work \\
21.93 & 20.42 $\pm$ 0.15 & $z'$ & LCO & This work \\
24.93 & 20.38 $\pm$ 0.09 & $z'$ & LCO & This work \\
29.71 & 20.75 $\pm$ 0.15 & $z'$ & LCO & This work \\
34.70 & 20.87 $\pm$ 0.15 & $z'$ & LCO & This work \\
38.33 & 21.12 $\pm$ 0.21 & $z'$ & NOT & This work \\
47.35 & 21.49 $\pm$ 0.14 & $z'$ & NOT & This work \\
50.35 & 21.9 $\pm$ 0.2 & $z'$ & NOT & This work \\

12.19 & 21.14 $\pm$ 0.04 & B & Zeiss-1000 & This work \\
13.20 & 21.06 $\pm$ 0.04 & B & Zeiss-1000 & This work \\
18.21 & 21.1 $\pm$ 0.1 & B & Zeiss-1000 & This work \\
19.19 & 21.06 $\pm$ 0.06 & B & Zeiss-1000 & This work \\
20.18 & 21.1 $\pm$ 0.05 & B & Zeiss-1000 & This work \\
22.17 & 21.47 $\pm$ 0.06 & B & Zeiss-1000 & This work \\
23.17 & 21.5 $\pm$ 0.06 & B & Zeiss-1000 & This work \\
12.22 & 20.16 $\pm$ 0.02 & V & Zeiss-1000 & This work \\
13.20 & 20.08 $\pm$ 0.03 & V & Zeiss-1000 & This work \\
18.24 & 20.17 $\pm$ 0.04 & V & Zeiss-1000 & This work \\
19.25 & 20.25 $\pm$ 0.03 & V & Zeiss-1000 & This work \\
20.26 & 20.24 $\pm$ 0.06 & V & Zeiss-1000 & This work \\
22.25 & 20.41 $\pm$ 0.04 & V & Zeiss-1000 & This work \\
23.25 & 20.5 $\pm$ 0.04 & V & Zeiss-1000 & This work \\
3.23 & 20.33 $\pm$ 0.14 & Rc & Zeiss-1000 & This work \\
6.38 & 20.21 $\pm$ 0.06 & Rc & NOT & This work \\
12.17 & 20.03 $\pm$ 0.02 & Rc & Zeiss-1000 & This work \\
13.18 & 19.98 $\pm$ 0.02 & Rc & Zeiss-1000 & This work \\
16.28 & 19.78 $\pm$ 0.04 & Rc & Zeiss-1000 & This work \\
18.18 & 19.94 $\pm$ 0.03 & Rc & Zeiss-1000 & This work \\
19.26 & 19.89 $\pm$ 0.02 & Rc & Zeiss-1000 & This work \\
20.25 & 19.96 $\pm$ 0.04 & Rc & Zeiss-1000 & This work \\
21.18 & 19.99 $\pm$ 0.04 & Rc & Zeiss-1000 & This work \\
22.23 & 20.03 $\pm$ 0.03 & Rc & Zeiss-1000 & This work \\
23.24 & 20.14 $\pm$ 0.03 & Rc & Zeiss-1000 & This work \\
30.15 & 20.5 $\pm$ 0.3 & Rc & Zeiss-1000 & This work \\
6.40 & 20.12 $\pm$ 0.1 & Ic & NOT & This work \\
12.24 & 19.5 $\pm$ 0.2 & Ic & Zeiss-1000 & This work \\
13.19 & 19.86 $\pm$ 0.06 & Ic & Zeiss-1000 & This work \\
18.26 & 19.93 $\pm$ 0.1 & Ic & Zeiss-1000 & This work \\
19.22 & 19.97 $\pm$ 0.06 & Ic & Zeiss-1000 & This work \\
20.21 & 20.05 $\pm$ 0.1 & Ic & Zeiss-1000 & This work \\
22.20 & 19.9 $\pm$ 0.07 & Ic & Zeiss-1000 & This work \\
23.20 & 20.08 $\pm$ 0.07 & Ic & Zeiss-1000 & This work \\
49.77 & 21.08 $\pm$ 0.12 & Ic & Keck I & This work \\
\end{longtable}
\clearpage

\begin{table*}
\centering
\small
\caption{Log of spectroscopy of SN\,2025kg.}\label{tab:opt_spec}
\begin{tabular}{ccccccc}
\toprule
UTC Date & $\Delta t$ & Range & Airmass  &  Total Exp. & Telescope  & Instrument \\
 & (day) &({\AA})&    & (s) & \\
\hline
 Jan. 18.85 & 10.35 & 4000--7600 & 1.6  &  2700 & GTC & OSIRIS   \\
 Jan. 18.88 & 10.36 & 4200--9000 & 1.6  &  4800 & NOT & ALFOSC   \\
 Jan. 19.86 & 11.34 & 4000--7000 & 1.3  &  1200 & SALT & RSS \\
 Jan. 22.07 & 13.55 & 4000--7000 & 1.1  &  2400 & SOAR & GHTS   \\
 Jan. 22.86 & 14.34 & 4000--8000 & 1.3  &  1200 & SALT & RSS   \\
 Jan. 23.08 & 14.56 & 4000--9000 & 1.2  &  1800 & Magellan & IMACS \\ 
 Jan. 24.13 & 15.61 & 4000--8000 & 1.3  &  1200 & SOAR   & GHTS \\
 Jan. 26.24 & 17.72 & 3000--8500 & 1.4 &  2700 & Keck~I   & LRIS\\
 Jan. 30.16 & 21.64 & 4000--9000 & 2.0 & 4860 & Shane & Kast \\
 Feb. 1.83 & 24.31 & 4000--9000 & 1.6  &  2400 & NOT  & ALFOSC \\
 Feb. 2.87 & 25.35 & 4000--9000 & 1.6  &  4800 & NOT   & ALFOSC\\
 Feb. 27.25 & 49.73 & 3000--8500 & 1.7 &  3600 & Keck~I   & LRIS\\
\bottomrule
\end{tabular}
\end{table*}
\clearpage

\begin{table*}
\centering
\small
\caption{Log of X-ray follow-up observations by EP/FXT and Swift/XRT}$^a$\label{tab:FXT_obs}
\begin{tabular}{ccccc}
\toprule
ObsID & Start time & End time & Exposure & Unabsorbed Flux$^{b}$\\
 & (UTC) &(UTC)& (s) & ($\rm erg\,s^{-1}\,cm^{-2}$) \\
\hline
EP/FXT \\
\hline
06800000359 & 2025-01-09T10:42:46 & 2025-01-09T13:09:02.017 & 6016 & $<\,7.5\times 10^{-15}$ \\
06800000366 & 2025-01-11T12:22:19 & 2025-01-11T13:12:41.755 & 3021 & \multirow{2}{*}{$<\,5.2\times 10^{-15}$} \\
06800000367 & 2025-01-11T15:34:34 & 2025-01-11T19:37:14.020 & 8423 &   \\
06800000376 & 2025-01-20T12:39:49 & 2025-01-20T18:18:03.017 & 12068 & $<\,5.1\times 10^{-15}$  \\
\hline
Swift/XRT \\ 
\hline
00019016001 & 2025-01-10T15:14:44 & 2025-01-11T02:19:17  & 2982 & $<\,8.0\times 10^{-14}$ \\
00019016002 & 2025-01-11T19:31:55 & 2025-01-12T06:30:22 & 6128 & $<\,2.5\times 10^{-14}$ \\
00019016003 & 2025-01-12T20:38:56 & 2025-01-13T10:30:58 & 6251 & $<\,2.5\times 10^{-14}$ \\
00019016004 & 2025-01-23T02:15:26 & 2025-01-23T06:17:30 & 1586 & \multirow{2}{*}{$<\,3.7\times 10^{-14}$} \\
00019016005 & 2025-01-23T06:55:56 & 2025-01-23T09:23:47 & 1746 &\\
\bottomrule
\end{tabular}
\begin{tablenotes}
\footnotesize
    \item[] $^a$We assume a $tbabs*powerlaw$ model in \texttt{Xspec} for the X-ray upper limit; the Galactic hydrogen column density is $1.6 \times 10^{20}\, \rm cm^{-2}$. The power-law index is assumed to be the same value ($-5.5$) as at late times of XRF\,060218. The intrinsic hydrogen absorption is not considered, as we do not have information on it. We stacked some observations close in time to get a deeper upper limit.
    \item[]  $^b$The unabsorbed flux upper limit at 0.5--10\,keV with 90\% confidence level.
\end{tablenotes}
\end{table*}
\clearpage

\begin{table*}
\centering
\caption{List of radio observations.}
\label{table:obs_radio}
\small
\begin{tabular}{cccc}
\toprule
UTC Date & Telescope &Frequency & Flux  Density ($3\sigma$)   \\
 (day) & &(GHz) & ($\mu$Jy)  \\
\midrule
Jan. 13 &MeerKAT& 3.0 &  $<$24 \\ 
Jan. 15 &VLA& 10 &  $<$16.5 \\ 
Jan. 16.44 &ATCA& 5.5 &  $<$24 \\ 
Jan. 16.44 &ATCA& 9.0 &  $<$18 \\ 
Feb. 01.48 &ATCA & 5.5 &  $<$33  \\ 
Feb. 01.48 &ATCA & 9.0 &  $<$30  \\
\bottomrule
\end{tabular}
\end{table*}

\clearpage
\end{document}